\documentclass[fleqn,usenatbib]{mnras}

\usepackage{newtxtext,newtxmath}

\usepackage[T1]{fontenc}

\DeclareRobustCommand{\VAN}[3]{#2}
\let\VANthebibliography\thebibliography
\def\thebibliography{\DeclareRobustCommand{\VAN}[3]{##3}\VANthebibliography}

\usepackage{graphicx}	
\usepackage{amsmath}	
\usepackage{caption}
\usepackage{cprotect}

\usepackage[dvipsnames]{xcolor}
\definecolor{tlmcolor}{HTML}{15998E}
\newcommand{\jr}{\color{black}}

\usepackage[normalem]{ulem}
\newcommand{\tlm}[2]{{\color{black}\sout{#1}#2}}

\title[Atmosphere emulation with neural networks]{Exoplanet atmosphere evolution: emulation with neural networks}

\author[J. G. Rogers et al.]{
James G. Rogers,$^{1}$\thanks{E-mail: james.rogers14@imperial.ac.uk}
Clàudia Janó Muñoz,$^{1,2,3}$
James E. Owen$^{1}$
and T. Lucas Makinen$^{1}$
\\
$^{1}$Astrophysics Group, Department of Physics, Imperial College London, Prince Consort Rd, London, SW7 2AZ, UK\\
$^{2}$Department of Physics and Astronomy, University of Manchester, Oxford Rd, Manchester M13 9PL, UK\\
$^{3}$Cavendish Laboratory, JJ Thomson Avenue, Cambridge CB3 0HE,
UK
}

\date{Accepted XXX. Received YYY; in original form ZZZ}

\pubyear{2022}

\begin{document}
\label{firstpage}
\pagerange{\pageref{firstpage}--\pageref{lastpage}}
\maketitle

\begin{abstract}
Atmospheric mass-loss is known to play a leading role in sculpting the demographics of small, close-in exoplanets. Knowledge of how such planets evolve allows one to ``rewind the clock'' to infer the conditions in which they formed. Here, we explore the relationship between a planet's core mass and its atmospheric mass after protoplanetary disc dispersal by exploiting XUV photoevaporation as an evolutionary process. Historically, this inference problem would be computationally infeasible due to the large number of planet models required; however, we use a novel atmospheric evolution emulator which utilises neural networks to provide three orders of magnitude in speedup. First, we provide proof-of-concept for this emulator on a real problem by inferring the initial atmospheric conditions of the TOI-270 multi-planet system. Using the emulator, we find near-indistinguishable results when compared to the original model. We then apply the emulator to the more complex inference problem, which aims to find the initial conditions for a sample of \textit{Kepler}, \textit{K2} and \textit{TESS} planets with well-constrained masses and radii. We demonstrate there is a relationship between core masses and the atmospheric mass they retain after disc dispersal. This trend is consistent with the `boil-off' scenario, in which close-in planets undergo dramatic atmospheric escape during disc dispersal. Thus, it appears the exoplanet population is consistent with the idea that close-in exoplanets initially acquired large massive atmospheres, the majority of which is lost during disc dispersal; before the final population is sculpted by atmospheric loss over 100~Myr to Gyr timescales.
\end{abstract}

\begin{keywords}
planets and satellites: atmospheres -
planets and satellites: physical evolution - planet star interactions
\end{keywords}



\section{Introduction}

The observed exoplanet population is dominated by planets with ages of 1-10~Gyr \citep[e.g.][]{McDonald2019,Berger2020b,Petigura2022}. Thus, it is distinctly separated in time from the formation process that happened early, in many cases within the first $\sim$10~Myr of the planet's life.  In order to connect observed exoplanets to their origins, we rely on the computation of evolutionary models to describe their possible histories. 

This issue is perhaps most pertinent for the small (1-4)~R$_\oplus$, close-in exoplanets (periods $\lesssim 100$~d) that are now thought to represent one of the dominant exoplanet populations \citep[e.g.][]{Howard2012,Fressin2013,Silburt2015,Mulders2018,Zink2019,Petigura2022}. In several cases, these planets are known to host H/He dominated atmospheres \citep[e.g.][]{Weiss2014,JontofHutter2016,Benneke2019}. Therefore, their proximity to the host star means they're vulnerable to atmospheric mass-loss, wherein the extreme irradiation drives powerful hydrodynamic outflows that cause the planet's atmosphere to lose mass \citep[e.g.][]{Baraffe2005,OwenJackson2012,Erkaev2016,OwenAlvarez2016,Kubyshkina2018}. It is now well established that atmospheric escape is capable of sculpting the close-in exoplanet population and is thought to play a key role in creating both the exoplanet desert and radius gap \citep[e.g.][]{Owen2013,LopezFortney2013,Owen2017,Ginzburg2018,OwenLai2018,Gupta2019,Wu2019,Owen2019,Gupta2020}. However, the formation scenario for this planetary population is uncertain, and strongly debated \citep[see recent review by][]{Bean2021}. These planets' bulk properties (mass and radius) vary significantly over their lifetimes due to a combination of cooling and mass-loss \citep[e.g.][]{Lopez2012,Owen2013}. Thus, computation of their evolution is critical to unravelling their formation since one can statistically constrain a planet's formation properties by determining which initial conditions can evolve into the planet we observe today \citep{Rogers2021}. 

However, atmospheric-loss-driven evolution is convergent: there are many initial planetary conditions that can evolve into an exoplanet with observationally indistinguishable bulk properties {\jr{\citep[e.g.][]{Owen2019,Owen2020,Kubyshkina2021}}}. This convergent evolution arises from both cooling and mass-loss. An initially hotter, higher entropy planet cools faster, meaning it reaches the same thermodynamic state as a planet that started cooler, with lower entropy. Similarly, planets with a more massive initial atmosphere are larger and, as such, can drive more powerful outflows, meaning it reaches the same atmospheric mass as a planet that started with a less massive atmosphere. 

Nevertheless, this convergent evolution does not mean a planet's initial conditions are completely lost. In fact, evolutionary modelling can provide important constraints that are inaccessible using only measurements of mass and radius for an evolved planet. Specifically, evolutionary modelling in both the photoevaporation and core-powered mass-loss scenario has indicated the core-composition of sub-Neptunes is an ``Earth-like'' iron-rock mixture \citep[e.g.][]{Owen2017,Wu2019,Gupta2019,Rogers2021}. Thus, evolutionary modelling allows one to break the well-known degeneracies in determining the compositions of small planets \citep[e.g.][]{Valencia2007,Rogers2010}. Furthermore, the link between planet radius, entropy and mass-loss means a planet's initial thermodynamic state is not completely lost. Specifically, a lower bound on a planet's initial cooling time (the Kelvin-Helmholtz timescale) can be determined because a planet with an even shorter cooling time would be larger and would have lost too much mass \citep[e.g.][]{Owen2020}. 

In this work, we aim to quantify the relation between core mass and the atmospheric mass retained after protoplanetary disc dispersal. This relation thus encodes the physics of gaseous core accretion during formation \citep[e.g.][]{Pollack1996,Lee2014,Massol2016} as well as atmospheric mass-loss during protoplanetary disc dispersal \citep[e.g.][]{Ikomi2012,Owen2016,Ginzburg2016}. The inference model involves finding the photoevaporative histories consistent with measured masses and radii for an ensemble of observed close-in exoplanets from \textit{Kepler}, \textit{K2} and \textit{TESS}. However, to fully and statistically characterise these plausible initial planetary conditions requires the computation of a large number of evolutionary models, and to do this at a population level is extremely computationally challenging. For example, to extract population level constraints on the exoplanet population at formation \citet{Rogers2021} evaluated $\sim 10^{10}$ planetary evolution models. As a result, this could only be applied over a narrow range of stellar mass, and expected correlations (for example, between core-mass and initial atmospheric mass) were not taken into account. 

The desired inference problem of this work will require the same, if not more evolutionary models to be evaluated. Therefore, we must consider a new, more computationally efficient approach. This is all the more pertinent as, in the end, one would like to use accurate evolution models that include, for example, a real equation of state, radiative transfer with atmosphere models and self-gravity. Fortunately, machine-learning provides an answer {\jr{and has been shown to be a powerful tool throughout a large range of scientific applications, particularly in the case of emulating the results complex numerical models at a fraction of the computational expense \citep[e.g.][]{Verrelst2015,Gilmer2017,Brehmer2018,Baydin2019,Tamayo2020,Himes2022}.}} Instead of solving the planetary structure and evolution equations for each planetary model one wishes to evolve, we can use a machine-learning model to emulate the planet's evolution. To demonstrate the feasibility of such an approach, here, we construct an exoplanet evolution emulator which is trained on semi-analytic models of XUV photoevaporation, testing it on the system TOI-270 \citep{VanEylen2021} before using it in the aforementioned inference problem. The success of this work highlights that, in the future, an emulator can be trained on considerably more computationally expensive but accurate models. 

In Section \ref{sec:Emulator}, we outline the emulator design and present benchmark tests. In Section \ref{sec:BHM}, we present the statistical inference model required to determine the correlation between core mass and atmospheric mass retention post disc dispersal, with results and discussion in Sections \ref{sec:results} and \ref{sec:discussion}. 

\section{Atmospheric evolution emulator} \label{sec:Emulator}

\begin{figure} 
	\includegraphics[width=0.95\columnwidth]{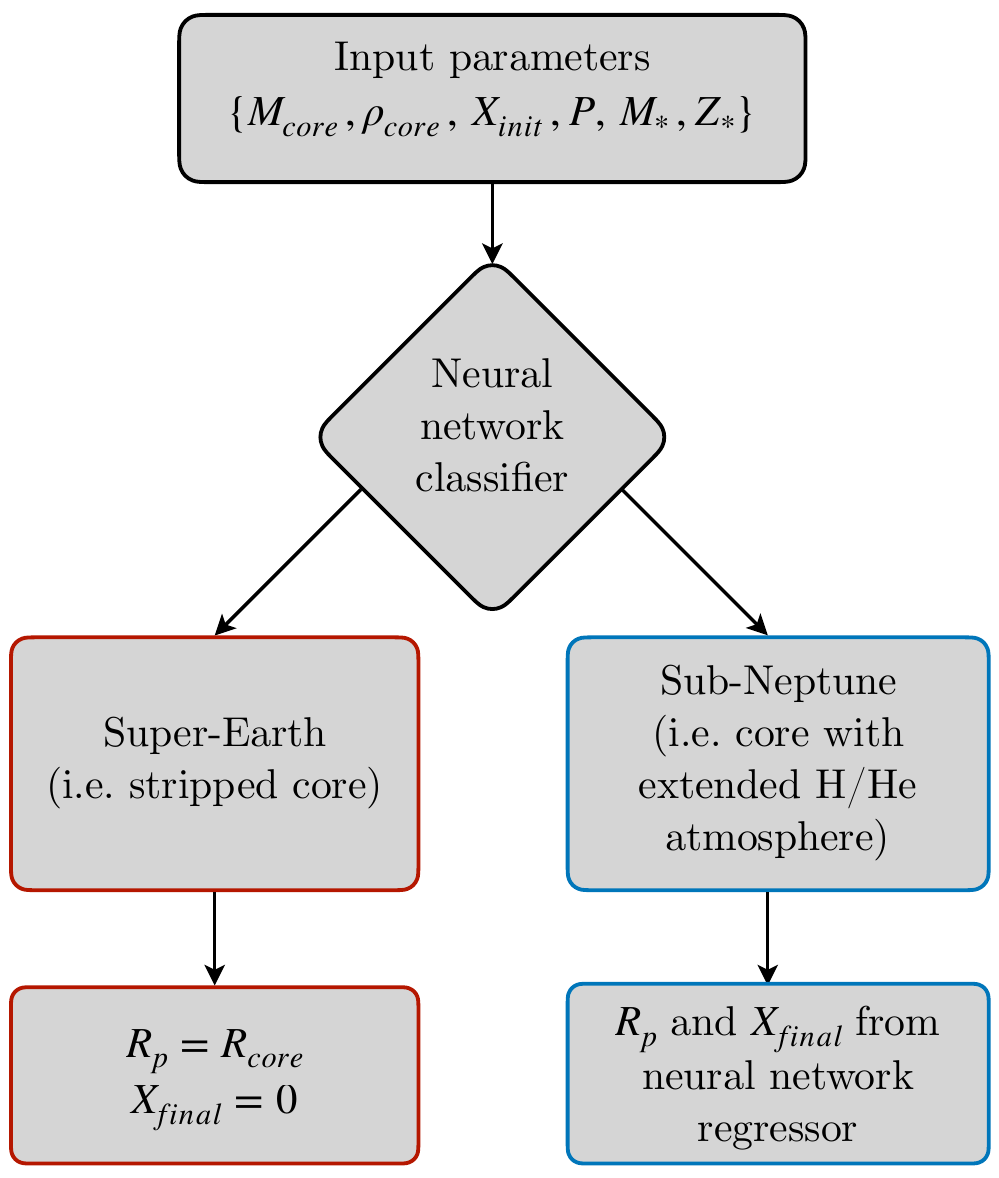}
    \caption{Flowchart to demonstrate the atmospheric evolution emulator. For a given set of parameters, an initial neural network classifier determines whether a planet will be stripped of its primordial H/He atmosphere. If so, the standard mass-radius relation of \protect{\citet{Fortney2007}} is used to provide the final radius. If the initial classifier predicts an atmosphere is retained, a neural network regressor predicts its final radius and atmospheric mass fraction.}
    \label{fig:Flowchart} 
\end{figure}

The purpose of an emulator is to reproduce the output of a physical model accurately whilst avoiding the computational expense of numerically solving the full problem (in our case, coupled ODEs). Since this is a proof of concept, we choose to train the evolution emulator with the semi-analytic model of the evolution of an exoplanet's H/He atmosphere from \citet{Owen2017,OwenEstrada2020}. This model calculates the evolution of a planet's photospheric radius and atmospheric mass-fraction as its primordial H/He atmosphere cools and experiences mass-loss due to XUV driven photoevaporation. Whilst we provide an overview here, we refer the interested reader to \citet{Owen2017,Rogers2021} and references therein for the technical details of this model.

The photoevaporation model of \citet{Owen2017,OwenEstrada2020} utilises an energy-limited mass-loss scheme, with atmospheric mass-loss rate, $\dot{M}_\text{atm}$, given as:
\begin{equation}
    \dot{M}_\text{atm} = \eta \frac{L_\text{XUV}}{4 a^2} \frac{R_\text{ph}^3}{ G M_\text{p}},
\end{equation}
where $\eta$ is the mass-loss efficiency, $a$ is the orbital semi-major axis, $R_\text{p}$ is the planetary radius and $M_\text{p}$ is the planetary mass. As in \citet{Rogers2021b}, the stellar X-ray/EUV luminosity, $L_\text{XUV}$, which is responsible for heating and ionising the H/He atmosphere is parameterised relative to the stellar bolometric luminosity, $L_\text{bol}$, by:
\begin{equation} \label{eq:LxLbol}
    \frac{L_\text{XUV}}{L_\text{bol}}=\begin{cases}
    \bigg( \frac{L_\text{XUV}}{L_\text{bol}} \bigg)_{\text{sat}} \; \bigg( \frac{M_*}{M_\odot} \bigg)^{-0.5} & \text{for } t < t_\text{sat}, \\
    \bigg( \frac{L_\text{XUV}}{L_\text{bol}} \bigg)_{\text{sat}} \; \bigg( \frac{M_*}{M_\odot} \bigg)^{-0.5} \; \bigg ( \frac{t}{t_\text{sat}} \bigg)^{-1-a_0}  & \text{for } t \geq t_\text{sat},
    \end{cases}
\end{equation}
where $a_0=0.5$ and $(L_\text{XUV} / L_\text{bol})_\text{sat} = 10^{-3.5}$, which is inspired by multiple observation works \citep[e.g.][]{Wright2011, Jackson2012, McDonald2019, Johnstone2021}. The saturation time $t_\text{sat}$ follows:
\begin{equation} \label{eq:tsat}
    t_\text{sat} = 10^{2} \; \bigg( \frac{M_*}{M_\odot} \bigg)^{-1.0} \;\; \text{Myr}.
\end{equation}
In order to calculate $L_\text{XUV}$, we use \textsc{MIST} stellar evolution tracks \citep{MIST-I2016, MIST-II2016} to accurately model the pre-main sequence and main-sequence evolution of the host star bolometric luminosity, $L_\text{bol}$, as a function of stellar mass, $M_*$, and metallicity, $Z_*$.

Whereas in previous iterations of this model, an analytic power law was used for the photoevaporative efficiency, $\eta$, in this work, we interpolate a grid of efficiencies calculated from the hydrodynamic models of \citet{OwenJackson2012}, which is also provided in the updated \verb|evapmass| code of \citet{OwenEstrada2020}. This approach provides physically accurate efficiencies as a function of planet mass and size. The original power law from \cite{Owen2017} was focused on explaining the sub-Neptune to super-Earth transition and neglected the efficiency drops for planets with low escape velocities due to advection \citep[e.g.][]{OwenJackson2012,Caldiroli2022}.

\begin{figure*} 
	\includegraphics[width=2.0\columnwidth]{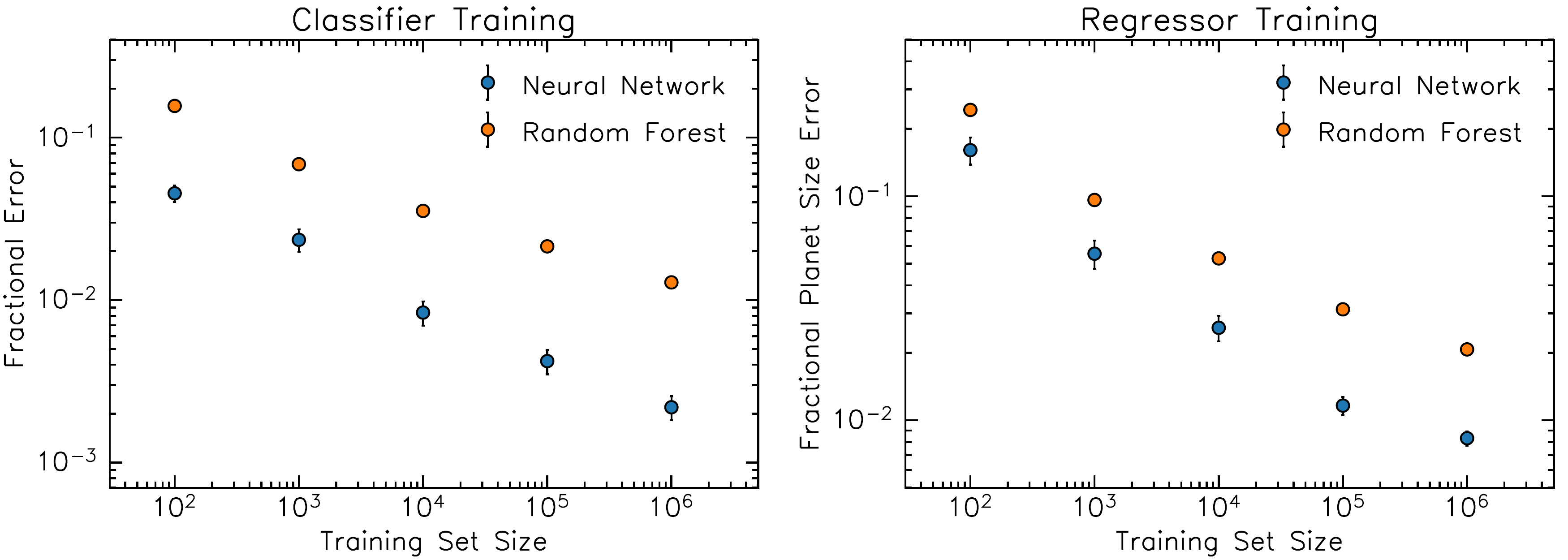}
    \cprotect\caption{The accuracy of the neural network emulator components is shown as a function of training set size compared to a random forest benchmark. Left: the classifier determines whether a planet is stripped of its H/He atmosphere or not, with accuracy measured by the fraction of incorrect classifications. Right: For those planets that maintain a significant atmosphere, the regressor calculates the final radius and atmospheric mass fraction of the planet. The accuracy is quantified in terms of the fractional difference between the planetary radius determined by the semi-analytic model and the emulator. Uncertainties are calculated by retraining the emulators 10 times and calculating the standard deviation in the accuracy. The neural network approach (blue) is compared to a similar emulator consisting of random forest algorithms (orange). Note also that typical uncertainties in planet radii are $\sim 5 \%$, which is surpassed by the neural network emulator for training sets of $\sim 10^4$ planets.}
    \label{fig:training} 
\end{figure*}

To emulate atmospheric evolution, we turn to supervised machine learning. The goal is to predict the photospheric radius and atmospheric mass fraction $X \equiv M_\text{atm} / M_\text{core}$ after billions of years of evolution without computing evolutionary tracks.  We wish to predict this ``final'' evolutionary state of a planet as a function of core mass $M_\text{core}$, initial atmospheric mass fraction $X_\text{init}$, orbital period $P$, host stellar mass $M_*$, host stellar metallicity $Z_*$ and core density $\rho_\text{core}$. We choose to interpret this latter variable in terms of a core composition $\tilde \rho$ according to the mass-radius relations of \citet{Fortney2007} in similar manners to \citet{OwenMorton2016,Rogers2021}. A composition of $\tilde \rho \leq 0$ signifies a ice-rock mixture, with $\tilde{\rho} = -1$ implying a $100\%$ ice core, ranging to $\tilde{\rho} = 0$ implying a $100\%$ rocky core. Similarly, $\tilde{\rho} \geq 0$ relates to a rock-iron mixture, with $\tilde{\rho}=1$ resulting in a $100\%$ iron core. This choice allows us to put plausible bounds on the core density;  however, it is important to note that the model constrains the bulk core density, not the composition. Thus, detailed constraints on the core composition require comparing the constrained bulk densities to detailed structure models.

\subsection{Emulator design} \label{sec:emulatorDesign}
The design of our emulator, as shown schematically in Figure \ref{fig:Flowchart}, is a two step-approach with all adopted algorithms taken from \verb|scikit-learn| \citep{scikit-learn}. In the photoevaporation model, the transition from a sub-Neptune (i.e. cores with extended H/He atmospheres) to a super-Earth (i.e. stripped cores) is quick since the mass-loss timescale drops rapidly for atmospheric mass fractions $\lesssim 1\%$, resulting in a runaway process \citep{Owen2013,LopezFortney2013,Owen2017,Mordasini2020}. Thus, the outcome of exoplanetary evolution is essentially bimodal. This bimodality motivates our two-step approach: we begin by implementing a multi-layer perceptron (MLP) neural network classifier to determine whether a given planet will evolve to become a super-Earth or a sub-Neptune. Since the planets that become super-Earths have a negligible atmospheric mass fraction, the final radius for these is simply given by the mass-radius relation of the stripped cores \citep[e.g.][]{Fortney2007}. For the sub-Neptunes, we employ a second machine learning algorithm, in this case, an MLP neural network regressor, to predict the sub-Neptune's final radius and atmospheric mass fraction. {\jr{To determine the architecture of both MLP networks, we utilised a grid-search over the number of hidden layers and nodes per layer, with a maximum of 200 nodes and 5 layers respectively. These maximum values were adopted since they provided an acceptable trade-off between training time and emulator accuracy. We found the optimum architecture for both networks included 4 hidden layers consisting of $(100, 100, 50, 15)$ nodes. We made use of a \verb|relu| activation function with an \verb|adam| \tlm{}{optimizer \citep{adam_optimizer},}  constant learning rates of $0.001$ \tlm{}{and batch size of $200$.}}} MLP algorithms were chosen over other ``classical'' machine-learning tools such as support vector machines, nearest neighbours, and random forests due to their superior performance in terms of accuracy and scalability with large data sets \citep[e.g.][]{goodfellow_deep_2016}. Specifically, as we discuss below, we found MLP algorithms to be more accurate and efficient than other methods. Thus, we were able to achieve the desired accuracy with reasonable amounts of computational resources.

To train the machine learning models, $10^6$ planets were simulated to $5$ Gyr with the semi-analytic photoevaporation model of \citet{Owen2017}, with the input parameters $\{ P, M_*, Z_*, M_\text{core}, \tilde{\rho}, X_\text{init}\}$ uniformly drawn such as to sample parameter space evenly. In this case, the bounds were as follows: $P \in [1,100]$ days, $M_* \in [0.35,1.5] M_\odot$, $Z_* \in [-1.0,0.5]$, $M_\text{core} \in [0.6,20.0]M_\oplus$, $\tilde{\rho} \in [-1.0,1.0]$ and $\log X_\text{init} \in [-4,0]$. {\jr{We normalise all variables to be in the range $[0,1]$ using a \verb|scikit-learn| \verb|MinMaxScaler| routine and produce a training/validation}} split of 75:25. \tlm{}{ Both networks were optimised on the training set, with final results reported for the validation data}. Figure \ref{fig:training} demonstrates the accuracy of the networks as a function of training data set size \tlm{}{on validation data}. \tlm{}{In the case of the classifier, the network is trained via gradient descent minimisation of cross-entropy loss}. The classifier accuracy in Figure \ref{fig:training} is represented via the fraction of cases in which a planet is misclassified as a super-Earth / sub-Neptune. For the case of the regressor, \tlm{}{the network is trained via gradient descent on the mean squared errors of} planetary radii and final atmospheric mass fraction between \tlm{}{true} and predicted \tlm{}{batch validation} data. The results are compared to an alternative machine learning approach using random forest algorithms instead of neural networks. {\jr{The random forests consist of 250 decision trees for the classifier and 90 decision trees for the regressor, which were chosen as a result of a parameter grid search}} \tlm{}{, trained on cross-entropy and mean square error losses, respectively.} One can see that the neural network approach outperforms the random forest in all cases. This behaviour is best explained in the context of network complexity. Neural networks are better suited to reproduce \tlm{}{arbitrary} non-linearities than decision trees, and arbitrarily deep neural network architectures, trained with gradient descent, have been shown to be universal function approximators \citep[e.g.][]{goodfellow_deep_2016}. 

Swapping a physical model for a machine learning emulator will substantially decrease the computational expense of calculations. In our case, it was found that for an equivalent set of planets, the emulator has a computational speedup factor of $\sim 1000$. However, the trade-off for an improved speed is the introduction of output error. As shown in Figure \ref{fig:training}, the accuracy of the classifier, if trained on $10^6$ planets, is $99.8\%$, with very occasional planets being misclassified as sub-Neptunes when they should be super-Earths. This latter point highlights a drawback of the semi-analytic model in which planets with low atmospheric mass fraction are inaccurately modelled, which we discuss in Section \ref{sec:emulatorDiscuss}. The RMS fractional error for the sub-Neptune regressor under similar training set sizes was $\sim 1\%$. This value is smaller than typical radii measurement uncertainties of $\sim 5\%$ \citep{Fulton2018, VanEylen2018}, {\jr{appropriate for current survey capabilities,}} implying that it is suitable for use in comparing a large number of evolutionary models to the observed bulk properties of exoplanets. {\jr{Future surveys, such as asteroseismic \textit{PLATO} measurements \citep{PLATO2014}, will likely reach this radii accuracy threshold of $\sim 1\%$ however, implying that further development may be required.}} Nonetheless, it is encouraging to note that the neural network emulator achieves this accuracy threshold with training set sizes as low as $\lesssim 10^4$. The RMS fractional error in the final atmospheric mass fraction was found to be $0.5\%$, and since the mass of a planet is dominated by the core, this introduces a negligible error in final mass. 

\begin{figure*} 
	\includegraphics[width=2.0\columnwidth]{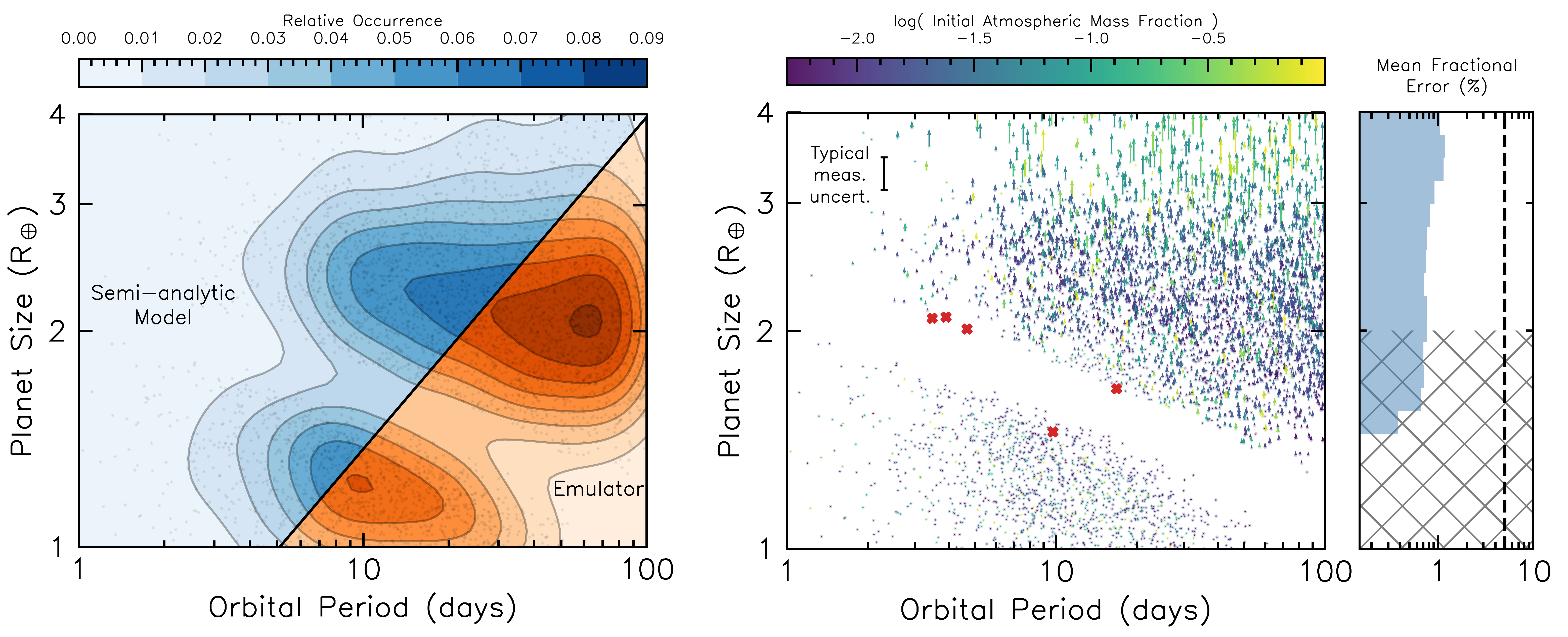}
    \cprotect\caption{Left: The radius gap computed with the semi-analytic model of XUV photoevaporation (orange) and with the evolution emulator (blue). Contours represent relative occurrence, and individual planets are plotted in black. Right: A population of planets are shown with their periods and final radii according to the semi-analytic model, with arrows pointing to this position as predicted with the evolution emulator. They hence represent the error between the physical model and the emulator for each planet in this test population. Note that the majority of planets do not have arrows since the error is negligible. Colours represent the initial atmospheric mass fraction. Whilst the average RMS error for the emulator is $1\%$, the right-hand histogram shows this error as a function of radius. {\jr{The grey-hatched region expresses the fact that whilst there is no error in planet size for super-Earths, this is because we inherently assume these planets have no atmosphere meaning that the mean fractional error is always zero if correctly classified.}} A typical radius measurement uncertainty is shown in black, taken from the value in the CKS catalogue \protect{\citep{Fulton2018}}.}
    \label{fig:comparison} 
\end{figure*}

In Figure \ref{fig:comparison}, we show {\jr{a population of test planets drawn from the underlying populations inferred in \citet{Rogers2021}}} within the period-radius plane evolved using the semi-analytic photoevaporation model in orange and evolution emulator in blue. One can see that the occurrence is very similar. We also show the errors of the emulator in Figure \ref{fig:comparison}, which are largest for larger planets with higher initial atmospheric mass-fraction. Despite constructing a training sample with uniformly drawn parameters, this does not evenly sample $R_\text{p}$ or $X_\text{final}$- space. Since the majority of planets evolve to smaller radii, this area of parameter space is more sparsely populated in the training data, and hence accuracy is lower here. Despite these errors being lower than those on real observations, we train our emulator on $10^6$ models to ensure the highest accuracy for all planet sizes. Finally, we show planets that are misidentified by the classifier as red crosses. Since the design of the emulator is for population-level inference analysis, these have a negligible  effect on any calculated constraints. {\jr{Nevertheless, if one considers the mass-loss timescale under XUV photoevaporation for misclassified planets, given by:
\begin{equation}
    t_{\dot{X}} = \frac{4GM_\text{c}^2 a X}{\eta L_\text{XUV} R_\text{ph}^3},
\end{equation}
then one finds that such planets are those losing mass on Gyr timescales. Since we evolve our training sample for $5$~Gyr, the emulator thus struggles to determine whether such planets will be stripped of their atmosphere in this length of time. We comment on this limitation in Section \ref{sec:emulatorDiscuss}.}}

\subsection{Initial Conditions of TOI-270} \label{sec:TOI-270}
To test the emulator {\jr{in a science-based case,}} we begin by inferring the initial conditions of TOI-270 b, c and d \citep{VanEylen2021}, which was discovered with \textit{TESS} \citep{Ricker2015,Gunther2019}. This archetypal system consists of an inner super-Earth (TOI-270 b) with an orbital period of 3.4 days and with observed mass and radius of $1.58\pm0.26 M_\oplus$ and $1.21\pm0.04 R_\oplus$. It also hosts two sub-Neptunes (TOI-270c and d) at 5.7 and 11.4 days, respectively, with masses of $6.15\pm0.37 M_\oplus$ and $4.78\pm0.43 M_\oplus$ and radii of $2.36\pm0.06 R_\oplus$ and $2.13\pm0.06 R_\oplus$. This architecture is highly indicative of mass-loss \citep[e.g.][]{OwenEstrada2020}. In a similar manner to the analysis of the Kepler-36 system in \citet{OwenMorton2016}, we set up a Bayesian model in which we aim to place constraints on the core composition $\tilde{\rho}$, core masses $M_\text{core}$ and initial atmospheric mass fractions $X_\text{init}$ required to evolve into the three planets that we observe today. Thus for our model parameters $\boldsymbol{\theta} = \{\tilde{\rho}, M_\text{core, b}, M_\text{core, c}, M_\text{core, d}, X_\text{init, b}, X_\text{init, c}, X_\text{init, d}\}$ and observed mass and radius data $\boldsymbol{D} = \{ R_\text{obs, b}, R_\text{obs, c}, M_\text{obs, b}, M_\text{obs, c}\}$, we may write from Bayes' law:
\begin{equation}
    P(\boldsymbol{\theta} | \boldsymbol{D}) \propto P(\boldsymbol{D} | \boldsymbol{\theta}) \; P(\boldsymbol{D}),
\end{equation}
where $P(\boldsymbol{\theta} | \boldsymbol{D})$ is the target posterior, $P(\boldsymbol{D})$ is the prior and assumed to be flat for core masses and core compositions and log-flat for initial atmospheric mass-fractions. Since all planets were formed in the same stellar environment, we follow \citet{OwenMorton2016} and assume the core composition is the same for all planets. The likelihood function $P(\boldsymbol{D} | \boldsymbol{\theta})$ is modelled as Gaussian:
\begin{equation}
   P(\boldsymbol{D} | \boldsymbol{\theta}) \propto \; \prod_i \exp \bigg (-\frac{(R_{\boldsymbol{\theta},i} - R_\text{obs,i})^2}{2\sigma^2_{R_i}} -\frac{(M_{\boldsymbol{\theta},i} - M_\text{obs,i})^2}{2\sigma^2_{M_i}} \bigg),
\end{equation}
where $i = \{b, c, d\}$ and $\sigma_{R_i}$ and $\sigma_{M_i}$ are the measurement uncertainties in radius and mass from transit and RV observations for each planet from \citet{VanEylen2021}. We evaluate the posterior by calculating the radius and mass of TOI-270 b, c and d for a given set of parameters $\boldsymbol{\theta}$ using the evolution emulator. In addition, we also take the measured stellar mass of $0.386 \pm0.008 M_\odot $, stellar metallicity of $-0.20 \pm 0.12$ and orbital periods from the planetary system, which are used as model input. To account for the measurement uncertainty in these quantities, we randomly redraw them from their associated errors every time the likelihood function is evaluated. To sample the posterior, we use the affine-invariant Monte Carlo Markov Chain (MCMC) \verb|emcee| \citep{ForemanMackey2014}. We run two independent chains that converge to the same results, each with 100 walkers and different initial conditions. {\jr{We justify chain convergence with the use of the Gelman-Rubin diagnostic \citep{gelman1992}, which yields a value of 1.0001, as well as the rank-normalised split-$\hat{R}$ diagnostic \citep{vehtari2021}, which yields a value of 1.001, which are both sufficiently close to unity to suggest MCMC chain convergence. The chain has an Effective Sample Size (ESS) $>10^3$, implying that the posteriors are sufficiently sampled.}}

\begin{figure*} 
	\includegraphics[width=1.9\columnwidth]{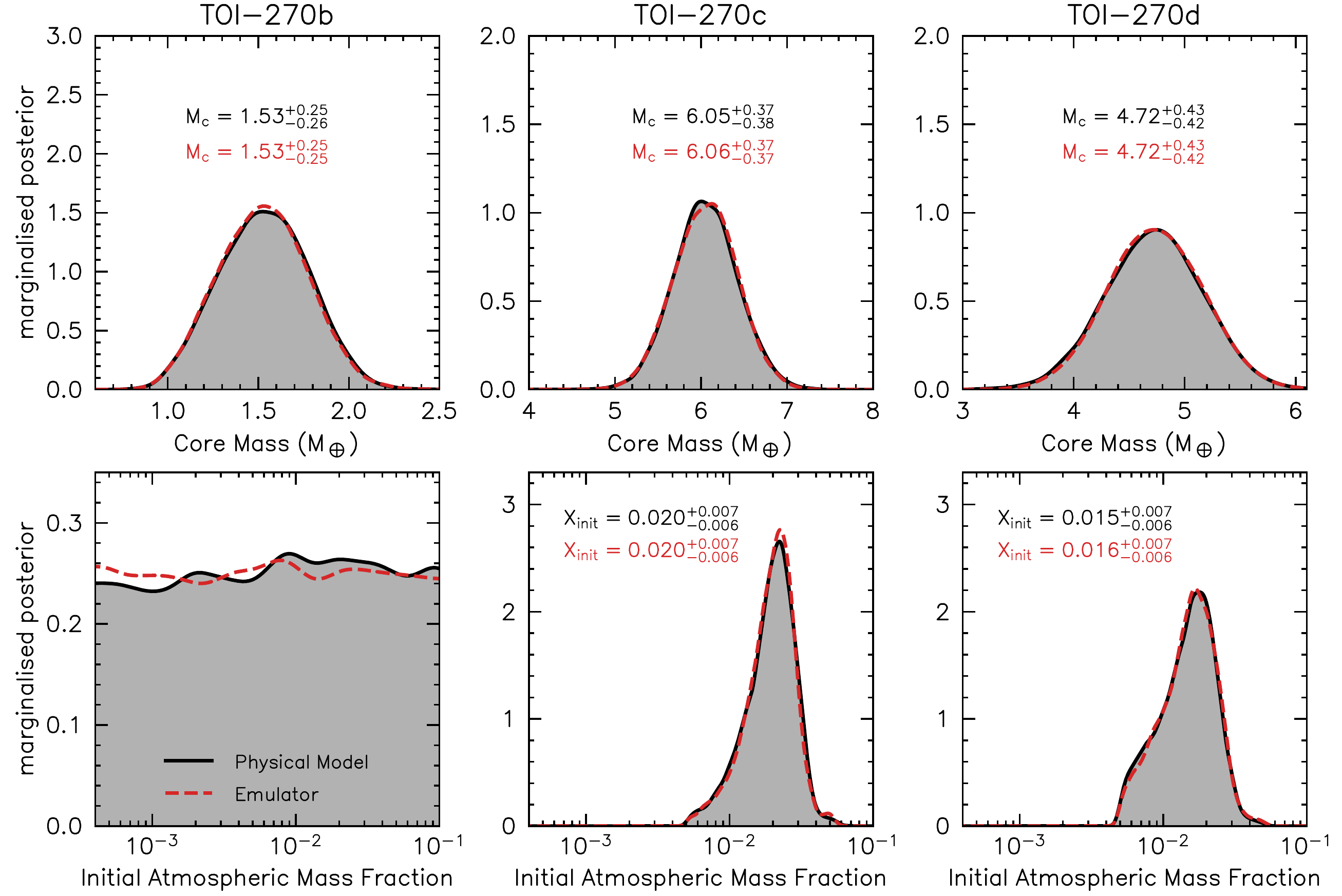}
    \cprotect\caption{Marginalised posteriors are shown for core masses (top row) and initial atmospheric mass fractions (bottom row) for TOI-270 b, c and d (left, middle and right-hand panels respectively). These are calculated with the semi-analytic model of XUV photoevaporation from \protect{\citet{Owen2017}} (black solid) and the machine learnt evolution emulator (red dashed), yielding near identical results.}
    \label{fig:TOI270posteriors} 
\end{figure*}

In Figure \ref{fig:TOI270posteriors}, we show marginalised posteriors for the core masses and initial atmospheric mass fractions for the TOI-270 system, as calculated by the evolution emulator. We also show posteriors which have been calculated with the semi-analytic model, which yields near identical results. Whilst the outer planets (TOI 270 c and d) are inferred to originally host atmospheric masses of $\sim 1-2\%$, the inner super-Earth (TOI 270 b) has a sufficiently small mass and close proximity to its host star to be completely stripped. The flat posterior on its initial atmospheric mass fraction indicates that the planet could have originally hosted any atmospheric mass and still been completely photoevaporated.

\begin{figure} 
	\includegraphics[width=0.95\columnwidth]{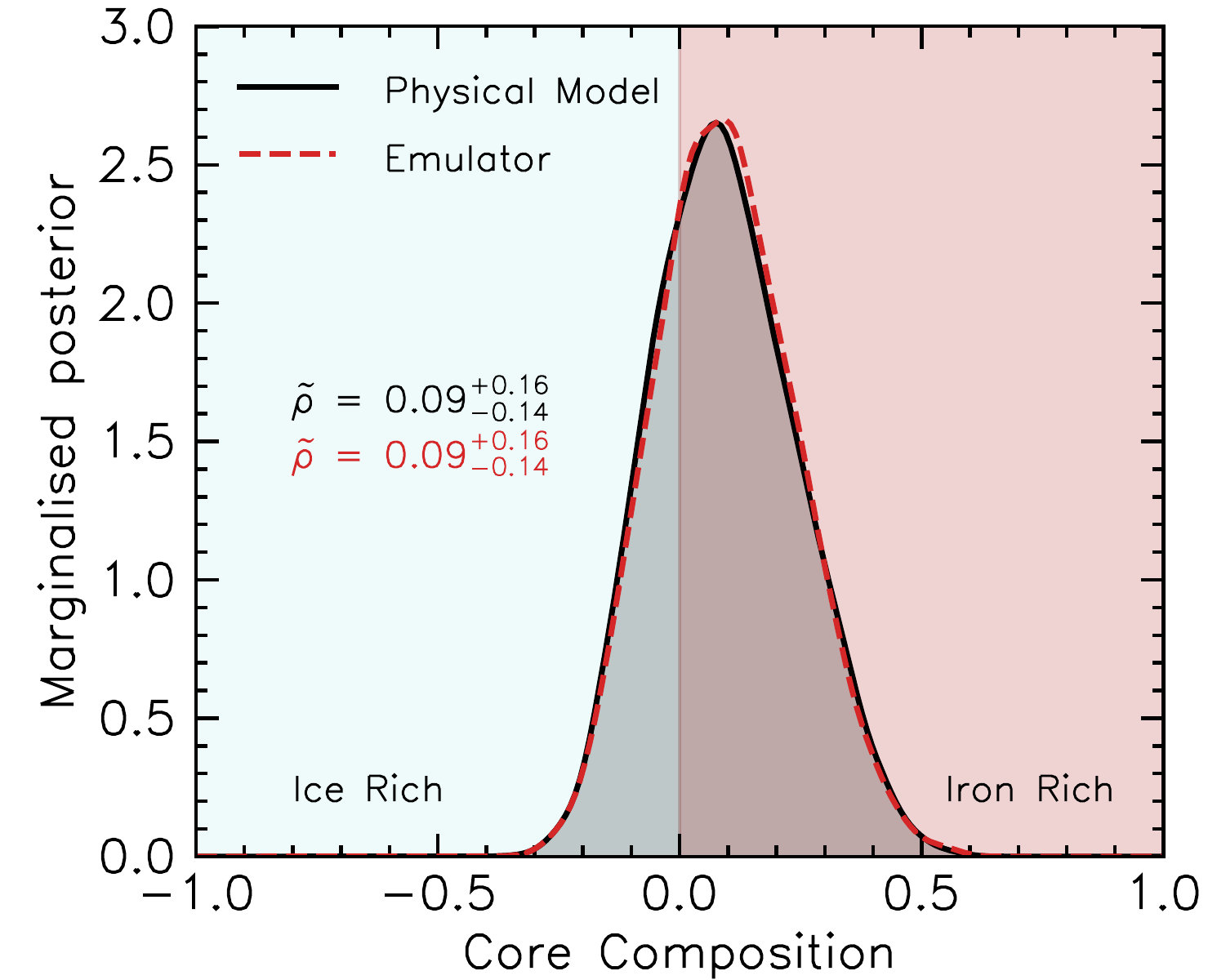}
    \cprotect\caption{Marginalised posterior for the core composition of TOI-270 planets. Negative values represent a composition consistent with an ice-rock mixture, whilst positive represents a rock-iron mixture. Similar to Figure \ref{fig:TOI270posteriors}, this is calculated with the semi-analytic model of XUV photoevaporation from \protect{\citet{Owen2017}} (black solid) and the machine learnt evolution emulator (red dashed), yielding near identical results.}
    \label{fig:TOI270composition} 
\end{figure}

In Figure \ref{fig:TOI270composition}, the marginalised posterior for core composition is shown for the TOI-270 planets, which is inferred to have an iron-mass fraction of $0.09^{+0.16}_{-0.14}$ and consistent with the population-wide inference analysis of \citet{Rogers2021}. This value is largely controlled by the measured mass and radius for planet b, for which its current mass and radii yield an iron-mass fraction of $\sim 0.1^{+0.15}_{-0.08}$. For reference, Earth has a composition in this parameterisation of $\tilde{\rho}_\oplus \approx 0.33$ \citep{Fortney2007}, which is consistent with TOI-270 to 2$\sigma$.

\section{Demographic Inference} \label{sec:BHM}

\begin{figure} 
	\includegraphics[width=0.95\columnwidth]{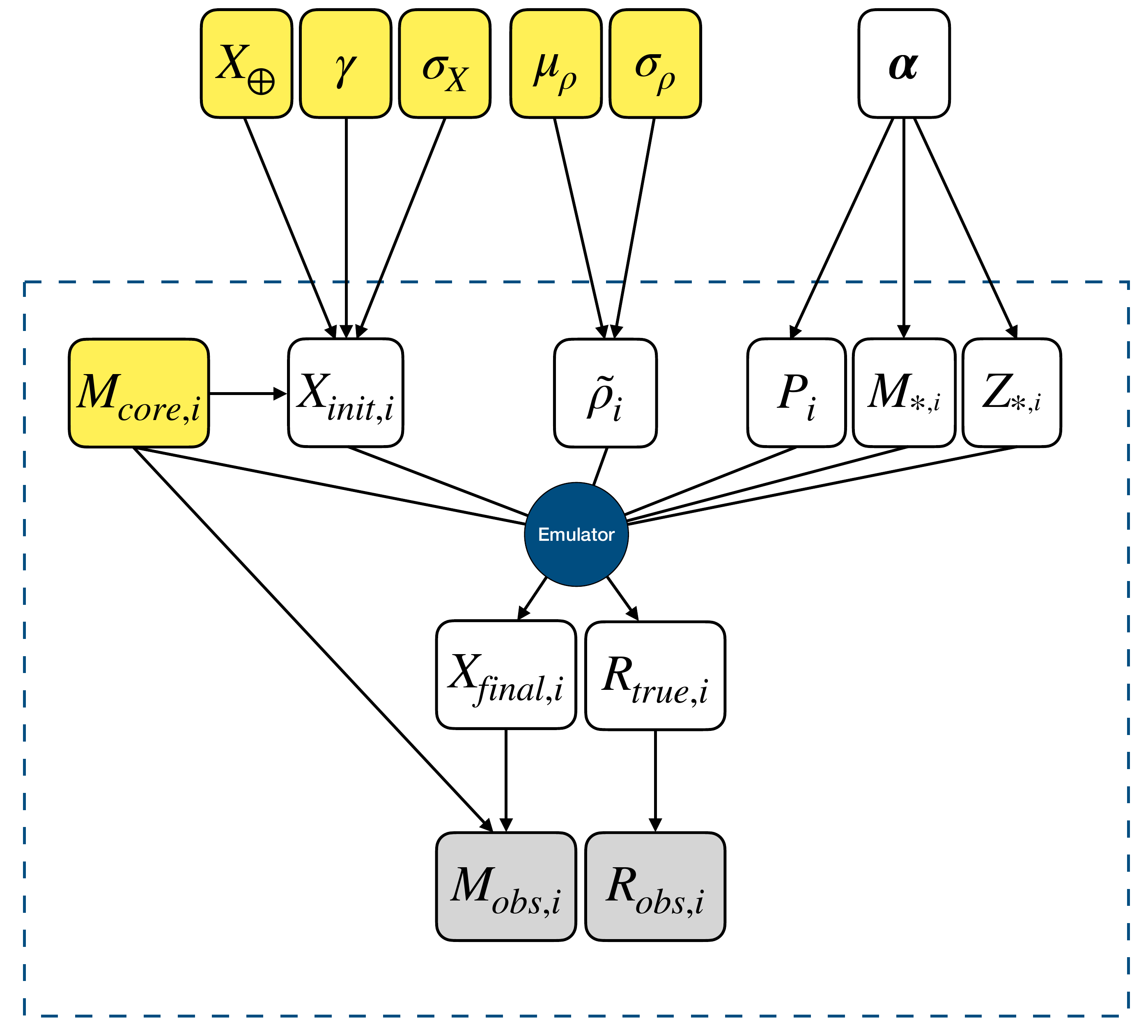}
    \cprotect\caption{Graphical representation of Bayesian hierarchical model for population of \textit{Kepler}, \textit{K2} and \textit{TESS} planets. Grey variables are directly observed with RV/TTV measurements for mass and transit observations for radii. White variables, meanwhile, are unobserved. Yellow variables are the parameters of interest in this problem and are constrained as part of the inference analysis. Variables outside of the dashed box are `hyperparameters' i.e. those that control population-level distributions, whilst each parameter inside the dashed box refers to the $i^\text{th}$ planet. Variable definitions are as follows: $X_\oplus$, $\gamma$ and $\sigma_X$ are hyperparameters that, along with core mass $M_\text{core}$, control the initial atmospheric mass fraction $X_\text{init}$ via Eq. \ref{eq:McoreVsXinit}; $\mu_\rho$ and $\sigma_\rho$ are hyperparameters for mean and standard deviation for Gaussian distribution that controls the core composition of planets $\tilde{\rho}$ via Eq. \ref{eq:BHMComps}; $\mathbf{\alpha}$ is a set of population-wide distribution hyperparameters that control the orbital period $P$, stellar mass $M_*$ and stellar metallicity $Z_*$ for each planet; the atmospheric evolution emulator (blue circle) determines the final atmospheric mass fraction $X_\text{final}$ and true radius $R_\text{true}$ of each planet; finally $M_\text{obs}$ and $R_\text{obs}$ are the observed mass and radius for each planet.} 
    \label{fig:BHMGraphical} 
\end{figure} 

In the previous demonstration of statistical inference on the TOI-270 system, we have shown that the emulator produces constraints that are in excellent agreement with the semi-analytic model it was trained on. However, the ultimate goal of the emulator is to exploit its computational speedup on large, population-level inference problems. 

We now apply the evolution emulator to a more sophisticated problem, which seeks to find the initial conditions for a population of exoplanets. In doing so, we seek to infer the relation between planetary core mass and the amount of atmospheric mass retained post-disc dispersal. This essentially sets the initial atmospheric conditions for photoevaporation for each planet. Chronologically speaking, planets will accrete H/He dominated material whilst immersed in their nascent disc, which have lifetimes of 3-10 Myrs \citep{Kenyon1995,Luhman2010,Ercolano2011,Luhman2010,Koepferl2013}. However, a large fraction of this mass has been shown to escape during disc dispersal on timescales of $10^5$ yrs \citep{Ikomi2012,Owen2016,Ginzburg2016}, through a process known as ``boil-off''. In this scenario, the nascent disc that once provided pressure support to a planet's newly formed atmosphere is removed faster than its own cooling timescale. As a result, the atmospheric material adiabatically expands and is removed from the planet's gravitational influence by the stellar bolometric luminosity via a Parker-type wind \citep{Parker1958}. This process halts once the planet's atmosphere has sufficiently cooled and contracted, typically such that its photospheric radius is $\sim 10\%$ the size of its Bondi radius \citep{Owen2016}.

For the proposed inference problem, we aim to place the first statistical constraint on how much atmospheric mass is retained after disc dispersal as a function of core mass. For this, we take a sample of exoplanets from the NASA Exoplanet Archive\footnote{Accessed on 09/08/2022, with all adopted quantities taken as default values from NASA Exoplanet Archive.} with measured masses and radii, with respective uncertainties of $\leq 20\%$ and $\leq 10\%$. Additional cuts were placed to restrict planet best-fit masses to $0.5 \leq M_\text{pl} \leq 20 M_\oplus$ and planet sizes to $0.6 \leq R_\text{pl} \leq 4 R_\oplus$. Additionally, stellar masses and metallicities were restricted such that $0.3 \leq M_* \leq 1.5 M_\odot$ and $-1.0 \leq Z_* \leq 0.5$. Finally, orbital periods were restricted to be $1 \leq P \leq 100$ days. These cuts were chosen so as to remain in the domain of high accuracy for photoevaporation models from \citet{Owen2017}. In particular, we avoid planets around late M-dwarfs ($M_* \leq 0.3 M_\odot$) since there is more uncertainty on the high-energy luminosity evolution (i.e. Eq. \ref{eq:LxLbol}) for such stars \citep[e.g.][]{Jackson2012,McDonald2019,Johnstone2021}. Note that this cut removes the Trappist-1 system from our planet sample \citep{Gillon2016}. After parameter cuts, 81 planets remain around 59 stars. The goal is to accurately reproduce these systems with an underlying physical model, utilising the computational speed of the atmospheric evolution emulator.

The inference problem takes the form of a Bayesian Hierarchical Model (BHM) and is outlined graphically in Figure \ref{fig:BHMGraphical}. The design of the inference model follows similar works from \citet{WolfgangLopez2015,Wolfgang2016}. For each planet, we draw a random core mass and draw its initial atmospheric mass fraction post-disc dispersal via the following prior distribution:
\begin{equation} \label{eq:McoreVsXinit}
	X_\text{init} \sim \mathcal{G} \bigg( \mu = X_\oplus \bigg( \frac{M_\text{core}}{M_\oplus} \bigg) ^\gamma, \sigma = \sigma_X \bigg).
\end{equation}
In other words, the prior on initial atmospheric mass fraction for each planet follows a power-law: $ X_\text{init} = X_\oplus (M_\text{core} / M_\oplus)^\gamma$ with a statistical scatter introduced via a Gaussian distribution $\mathcal{G}$ with a standard deviation of $\sigma_X$. Thus, we are aiming to infer the values of $X_\oplus$, being the mean initial atmospheric mass fraction for an Earth-mass core; $\gamma$, being the power-law index; and $\sigma_X$, being the intrinsic scatter involved in the associated atmospheric accretion and escape processes. In a similar vein, we also aim to infer the core compositions of planets via a separate Gaussian prior:
\begin{equation} \label{eq:BHMComps}
	\tilde{\rho} \sim \mathcal{G}( \mu =\mu_\rho, \sigma = \sigma_\rho)
\end{equation}
{\jr{where we aim to infer the parameters $\mu_\rho$ and $\sigma_\rho$. Since these parameters and those that describe initial atmospheric mass fraction are population-level parameters, they are labelled as hyperparameters in the nomenclature of Bayesian hierarchical models.}}

{\jr{Note that since we are attempting to infer population-level distributions for core composition and initial atmospheric mass fraction, it is important to perform this inference on all planetary systems in our sample simultaneously, as opposed to performing individual inferences for each system. For example, performing this analysis on a single sub-Neptune in a given system would yield uninformative constraints due to the known degeneracy in planet composition with measured mass and radii \citep[e.g.][]{Rogers2010}. We can break this degeneracy in a multi-planet system which hosts both super-Earths and sub-Neptunes, such as the case of TOI-270 from Section \ref{sec:TOI-270}, since there are only certain combinations of initial atmospheric mass-fractions and core masses that reproduce the planet observations (assuming they have the same core composition). Analogously, for our population level inference presented here, we must perform the analysis on a population of planets consisting of super-Earths and sub-Neptunes under the assumption that they follow population-wide distributions.}} 

Whereas in previous analyses that use similar statistical frameworks to infer population-level hyperparameters controlling the distribution of core masses \citep{Wolfgang2016,Rogers2021}, we choose to instead infer the core mass for every planet as an individual parameter. In total, this inference model thus consists of 5 hyperparameters and 81 parameters (being the core masses for each planet), meaning the posterior is sampled in 86-dimensional parameter space. Without the computational speed of the atmospheric evolution emulator, this would be a truly futile endeavour. Nonetheless, we choose not to infer individual parameters for core composition and initial atmospheric mass fraction since this would result in an impractically large inference problem.

For each planet: stellar masses, stellar metallicities and orbital periods are drawn from within their associated measured values. Similar to the analysis of TOI-270 in Section \ref{sec:TOI-270}, we assume that planets within a given system have the same core composition. The likelihood function is calculated as follows
\begin{equation}
   P(\boldsymbol{D} | \boldsymbol{\theta}) \propto \; \prod_i \exp \bigg ( w_i \bigg( -\frac{(R_{\boldsymbol{\theta},i} - R_\text{obs,i})^2}{2\sigma^2_{R_i}} -\frac{(M_{\boldsymbol{\theta},i} - M_\text{obs,i})^2}{2\sigma^2_{M_i}} \bigg) \bigg), 
\end{equation}
where $i$ denotes an individual planet from the sample. Here, $w_i$ is a weighting that is required to account for an intrinsic bias in mass measurements that favours larger mass planets. This results in a sample consisting of far more sub-Neptunes than super-Earths, which is not representative of the underlying distribution. In order to infer hyper-parameters that are indeed representative, we weight by a completeness corrected radii distribution of Kepler planets. Hence this should make our estimate of atmospheric mass as a function of core mass more representative of the underlying Kepler sample. To calculate this weight, we determine a probability density function (PDF) of planetary radii for the adopted sample, labelled $\mathcal{P}_\text{obs}(R / R_\oplus)$. Additionally, a second PDF for planet radii is taken from the inference analysis of \cite{Rogers2021}, which represents the underlying population of Kepler planets, i.e. an unbiased distribution of planet radii, which is labelled as $\mathcal{P}_\text{pop}(R / R_\oplus)$. The weight for a given planet, $w_i$, is the ratio of these two distributions, i.e. $w_i = \mathcal{P}_\text{pop}(R_i / R_\oplus) \, / \, \mathcal{P}_\text{obs}(R_i / R_\oplus)$. However, when calculating this numerically, the weight diverges when $\mathcal{P}_\text{obs}$ tends to zero in regions of no occurrence. Therefore, we introduce a small number, $\epsilon$, to avoid this problem:
\begin{equation}
	w_i = \frac{\mathcal{P}_\text{pop}(R_i / R_\oplus) + \epsilon}{\mathcal{P}_\text{obs}(R_i / R_\oplus) + \epsilon}.
\end{equation}
Since the choice of $\epsilon$ is arbitrary, we choose its value to be $\epsilon=1$. One can see that in the limit that $\mathcal{P}_\text{pop}$ tends to $\mathcal{P}_\text{obs}$, the weight is unity for all planet sizes. In our case, there are more super-Earths in $\mathcal{P}_\text{pop}$ than in $\mathcal{P}_\text{obs}$. Hence they are given a larger weight, resulting in a likelihood that samples a fair approximation of the underlying distribution of \textit{Kepler} planets. 

{\jr{All hyperparameters and core mass parameters are given flat priors (except $X_\oplus$ which is given a log-flat prior) and are thus uninformative on the inference results.}} Posteriors are sampled with the affine invariant Monte Carlo Markov Chain of \citet{GoodmanWeare2010} from \verb|emcee| \citep{ForemanMackey2014}. We run two independent chains that converge to the same results, each with 1000 walkers and different initial conditions. {\jr{As with the inference analysis of Section \ref{sec:TOI-270}, we justify chain convergence with the use of the Gelman-Rubin diagnostic \citep{gelman1992} and rank-normalised split-$\hat{R}$ diagnostic \citep{vehtari2021} which yield values of 1.0002 and 1.01 respectively, which are both sufficiently close to unity to suggest MCMC chain convergence. The chain has an Effective Sample Size (ESS) $>10^3$, implying that the posteriors are sufficiently sampled.}}

\section{Results} \label{sec:results}

\begin{figure*} 
	\includegraphics[width=1.9\columnwidth]{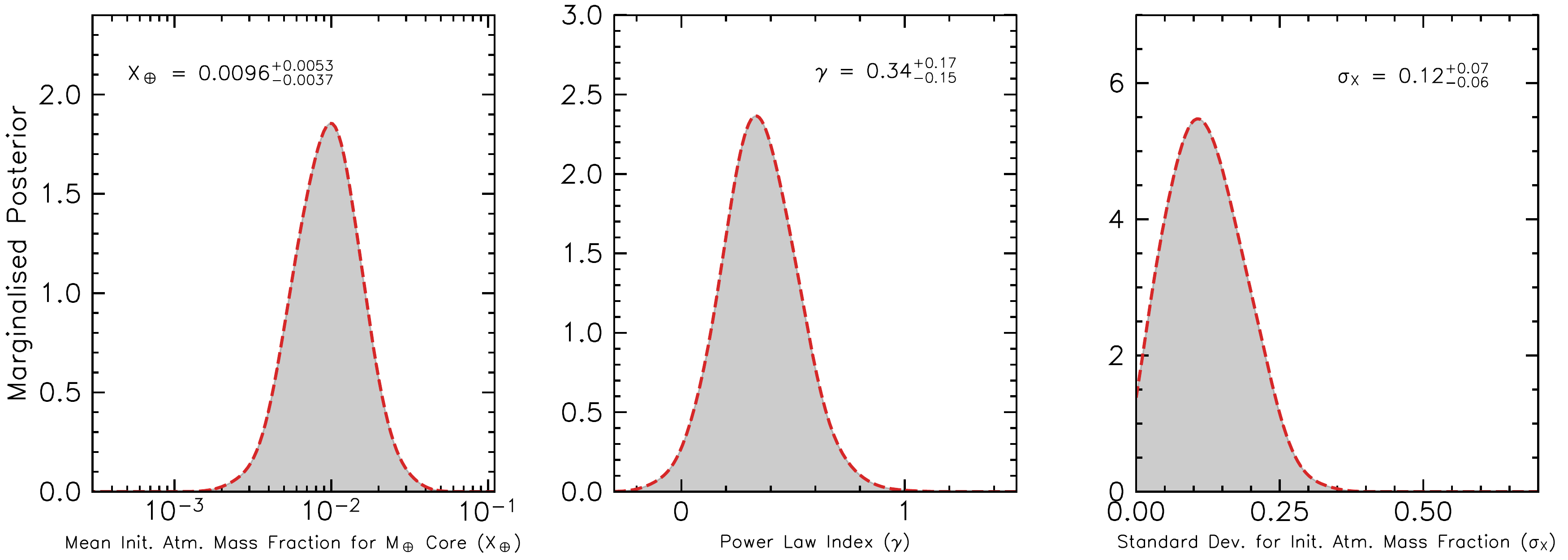}
    \cprotect\caption{Marginalised posteriors for parameters controlling the initial atmospheric mass fraction post disc dispersal for our sample of \textit{Kepler}, \textit{K2} and \textit{TESS} planets, which follows a Gaussian distribution of the form: $X_\text{init} \sim \mathcal{G} ( \mu=X_\oplus (M_\text{core} / M_\oplus)^\gamma, \sigma=\sigma_X)$.}
    \label{fig:BHM_hyperX} 
\end{figure*}

\begin{figure*} 
	\includegraphics[width=1.9\columnwidth]{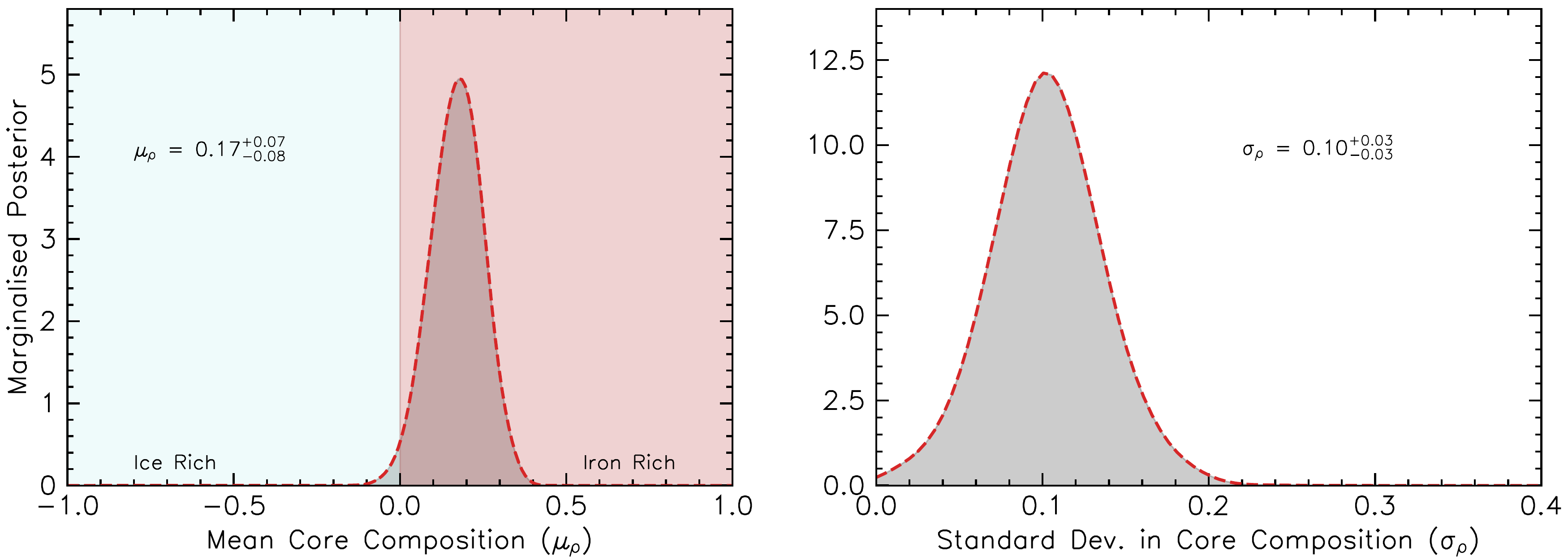}
    \cprotect\caption{Marginalised posteriors for the core composition of the adopted sample of \textit{Kepler}, \textit{K2} and \textit{TESS} planets, which is modelled as a Gaussian distribution with mean $\mu_\rho$ and standard deviation $\sigma_\rho$. Left: For the mean core composition, negative values represent a composition consistent with an ice-rock mixture, whilst positive represents a rock-iron mixture. Right: the standard deviation represents the spread in core compositions for the planet sample.}
    \label{fig:BHM_hyperComp} 
\end{figure*}

Posteriors for the population level hyperparameters are shown in Figures \ref{fig:BHM_hyperX} and \ref{fig:BHM_hyperComp}. For the hyperparameters controlling the relation between core mass and initial atmospheric mass fraction post disc dispersal (see Eq. \ref{eq:McoreVsXinit}), we infer $X_\oplus = 0.0096^{+0.0053}_{-0.0037}$, implying that an Earth-mass core will retain $\sim 1\%$ atmospheric mass fraction once its protoplanetary disc has dispersed. Additionally, we find that the power-law index is $\gamma=0.34^{+0.17}_{-0.15}$. Finally, we find the standard deviation in this atmospheric mass retention is $\sigma_X = 0.12^{+0.07}_{-0.06}$, implying there is an astrophysical scatter involved of $\sim 0.12$ dex.

For hyperparameters controlling the Gaussian distribution of planetary core compositions (see Eq. \ref{eq:BHMComps}), we find a mean of $\mu_\rho = 0.17^{+0.07}_{-0.08}$, which is consistent with the results for TOI-270 in Figure \ref{fig:TOI270composition} and the inference analysis of \citet{Rogers2021}. We find the standard deviation in compositions to be $\sigma_\rho = 0.10^{+0.03}_{-0.03}$, implying a $\sim 0.1$ dex spread in core compositions for the chosen sample of planets. 

\begin{figure} 
	\includegraphics[width=0.95\columnwidth]{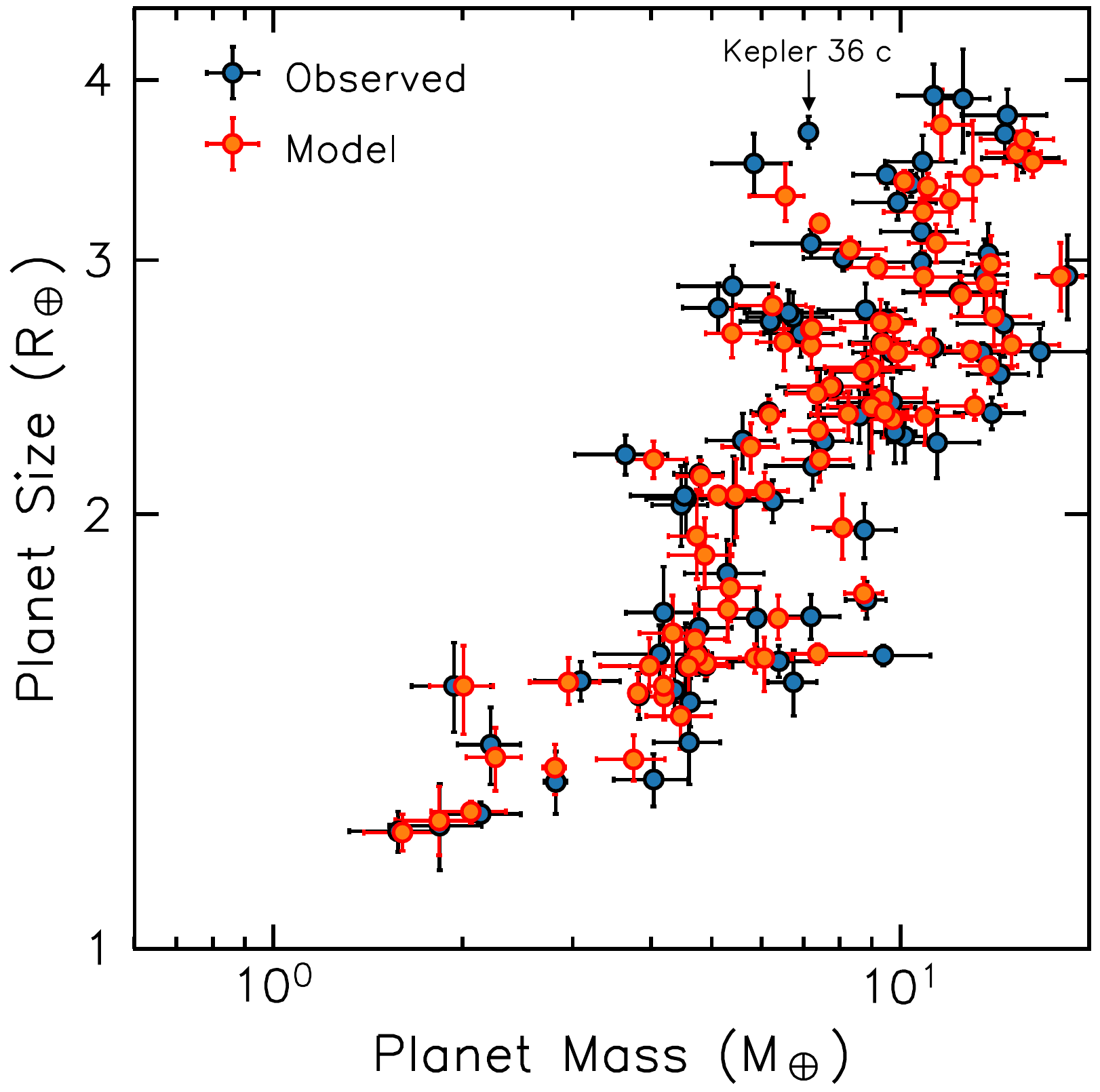}
    \cprotect\caption{Planet masses and radii are shown in blue for the selected planet sample from \textit{Kepler}, \textit{K2} and \textit{TESS}, with error bars from their associated measurement uncertainty. The suitability of the model presented in Figure \ref{fig:BHMGraphical} is demonstrated with latent variables in orange, i.e. the values of planet mass and radii for each planet as calculated through the Bayesian Hierarchical Model. Error bars are representative of a $1\sigma$ {\jr{credible region, calculated by integrating the MCMC posterior distributions. Kepler 36 c is the only system that is not consistent with our statistical model.}}}
    \label{fig:Latent} 
\end{figure}

\begin{figure}
	\includegraphics[width=1.0\columnwidth]{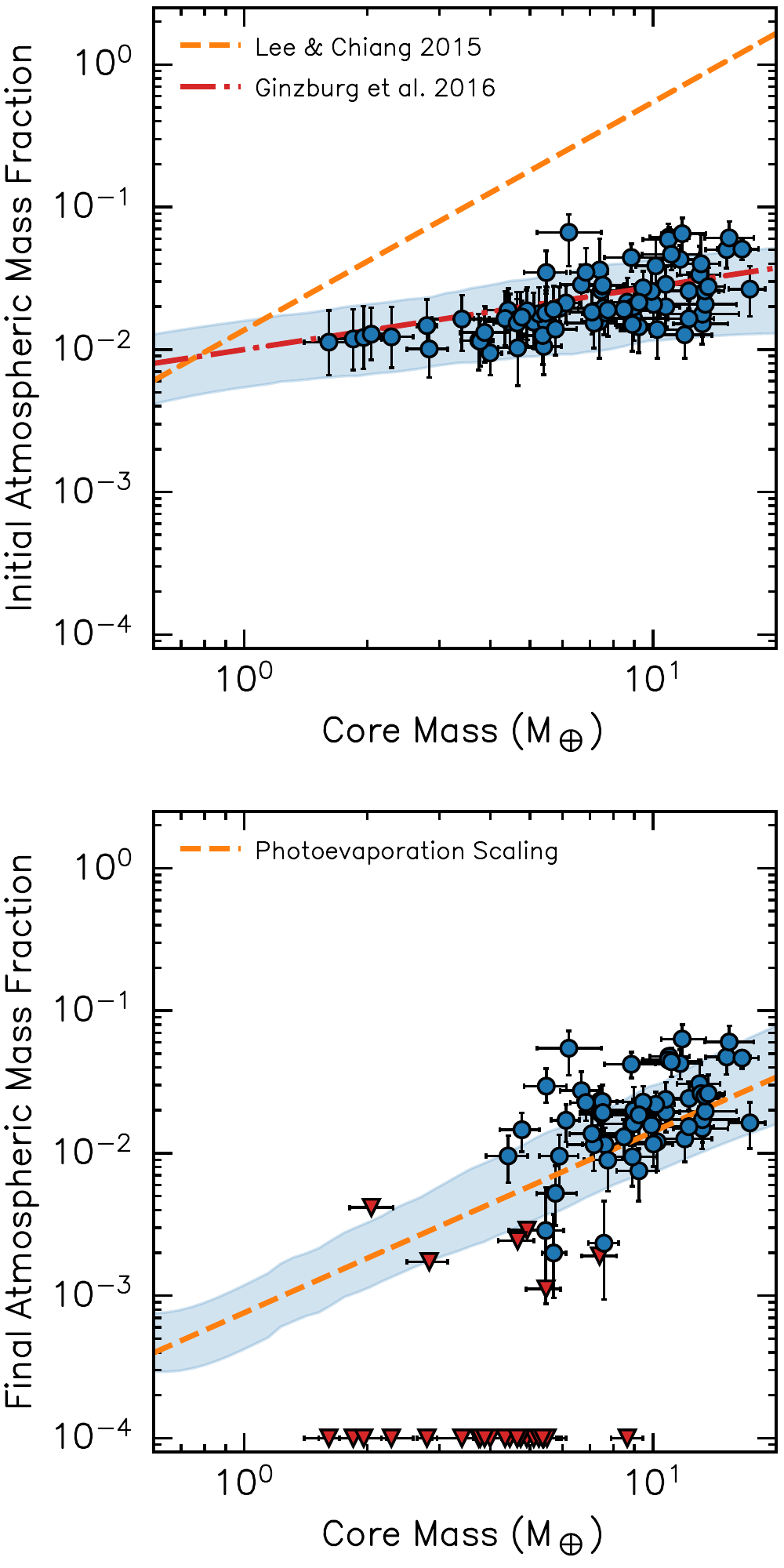}
    \cprotect\caption{Top: the distribution of initial atmospheric mass fractions post disc dispersal for the planet sample is shown as a function of core mass, constrained via Eq. \ref{eq:McoreVsXinit}. The blue-shaded region is representative of a $1\sigma$ {\jr{credible region for the form of this relation, calculated by integrating the MCMC posterior distributions.}} The analytic core accretion model from \citet{Lee2015} is shown in dashed orange. The analytic `spontaneous mass-loss' model of \citet{Ginzburg2016} \citep[also referred to as `boil-off', e.g.][]{Owen2016} is shown in red dot-dashed, which represents a prediction of atmospheric retention post disc-dispersal. Bottom: The final atmospheric mass fraction distribution of the chosen planet sample. Planets with a final atmospheric mass fraction consistent with $10^{-4}$ are inferred to have been stripped of their primordial H/He atmosphere and are represented as upper limits on the final atmospheric mass fraction (red triangles). An analytic scaling relation for the atmospheric mass fractions of sub-Neptunes (see Eq. \ref{eq:PEscaling}) is shown in dashed orange.}
    \label{fig:XinitXfinal} 
\end{figure}

To assess the suitability of these inferred parameters, Figure \ref{fig:Latent} shows the sample of planets from \textit{Kepler}, \textit{K2} and \textit{TESS} in blue. In orange, we show the set of latent variables i.e. values of planet mass and radii for each planet as calculated through the Bayesian hierarchical model. Apart from one exception, all modelled planets are consistent with the observed values, implying excellent agreement between data and model. Kepler 36 c, however, appears as an outlier in Figure \ref{fig:Latent}. We discuss Kepler 36 further in Section \ref{sec:Kepler36c}

In Figure \ref{fig:XinitXfinal}, we present the inferred core masses with initial and final atmospheric mass fractions for the planet sample. Note we provide these inferred values for all planets in Table \ref{tab:models}. In the top panel, the initial values are shown alongside the inferred form of Eq. \ref{eq:McoreVsXinit} as a blue {\jr{credible region}}, which relates core mass to this initial atmospheric mass. All uncertainties represent $1\sigma$ {\jr{credible regions}} and are calculated by integrating the MCMC posterior distributions. In addition, analytic models are shown from \citet{Lee2015}, which predicts the atmospheric mass attained during gaseous core accretion, and from \citet{Ginzburg2016}, which predicts the atmospheric mass retained post disc-dispersal and the subsequent boil-off process (referred to in this work as `spontaneous mass-loss'). One can see that the inferred relation between core mass and initial atmospheric mass fraction is consistent with that of \citet{Ginzburg2016}, suggesting that small, close-in exoplanets do indeed undergo a boil-off process after core accretion and during disc dispersal. We comment on these findings in Section \ref{sec:BHMDiscuss}. 

Finally, in the bottom panel of Figure \ref{fig:XinitXfinal}, the final atmospheric mass fractions are shown as a function of core mass with uncertainties calculated in the same way. Planets with a final atmospheric mass fraction consistent with $10^{-4}$ are shown with upper limits (red triangles) and are inferred to have been stripped of their primordial H/He atmosphere within their lifetime. Such planets tend to be lower in mass since they have a lower gravitational influence and are more vulnerable to photoevaporation. We fit the final atmospheric mass fractions for planets that maintain a significant atmosphere (i.e. blue circles) with a power law of the form Eq. \ref{eq:McoreVsXinit}, which is shown as a blue {\jr{credible region}} representative of $1\sigma$ uncertainties. One can analytically predict the form of this relation, with the use of Eq. 14 from \citet{Owen2017}, which states that the atmospheric mass fraction of a planet for a given size and core mass is:
\begin{equation}
    X \propto M_\text{c}^{0.17} \, \bigg( \frac{\Delta R}{R_\text{c}} \bigg )^{1.31},
\end{equation}
where $R_\text{c}$ is the core radius and $\Delta R = R_\text{p} - R_\text{c}$ represents the height of the planet's optically thick atmosphere. In addition, Eq. 20 from \citet{Owen2017} relates the photoevaporative mass-loss timescale to these same variables:
\begin{equation}
    t_X \propto M_\text{c}^{1.42} \, \bigg( \frac{\Delta R}{R_\text{c}} \bigg )^{1.69}.
\end{equation}
Note that these relations from \citet{Owen2017} assume an adiabatic index of $\gamma = 5/3$ and an opacity scaling relation of $\kappa \propto P^{\alpha} T^{\beta}$, where $\kappa$, $P$ and $T$ are the opacity, pressure and temperature respectively, with $\alpha=0.68$ and $\beta=0.45$ chosen from  \citet{RogersSeager2010}. This is appropriate for a highly irradiated low-mass planet with a solar metallicity H/He atmosphere. Since all sub-Neptunes (i.e. planets that have retained a significant atmospheric mass fraction) will have been photoevaporated for the same length of time, we can state that the mass-loss timescale is constant for such planets. This then allows one to eliminate $\Delta R / R_\text{c}$ to find:
\begin{equation} \label{eq:PEscaling}
    X_\text{final} \propto M_\text{c}^{1.27} 
\end{equation}
This approximate power law is shown in the bottom panel of Figure \ref{fig:XinitXfinal}, which also shows good agreement with the inferred scaling relation.

\section{Discussion} \label{sec:discussion}
Our inference analysis has demonstrated the capability of machine-learnt emulators in exoplanet evolution studies. We began by training a two-step neural network emulator on the semi-analytic models of XUV photoevaporation from \citet{Owen2017, OwenEstrada2020}. As a proof of concept, we applied the emulator to the archetypal system TOI-270 \citep{VanEylen2021}, consisting of three planets that straddle the radius gap. We used the emulator to find the initial conditions required to evolve into the planets that we observe today. We find that the emulator places constraints on core mass, initial atmospheric mass fraction and core composition that are near-indistinguishable from those calculated by the semi-analytic model that it was trained on. 

In light of this, we then applied the emulator to a far more complex inference problem, which sought to infer the initial conditions for a sample of 81 \textit{Kepler}, \textit{K2} and \textit{TESS} planets with well-constrained masses and radii. Consistent with the result for TOI-270, we find that the core compositions of such planets have an iron-mass fraction of $0.17^{+0.07}_{-0.08}$. In addition, we present the inferred core masses, initial atmospheric mass fractions and final atmospheric mass fractions for all planets in Table \ref{tab:models}.

\subsection{The initial conditions of TOI-270} \label{sec:TOI-270Discussion}
Using both the machine-learnt evolution emulator and the semi-analytic model from which it was trained, we find that the core masses for TOI-270 b, c and d are $1.53^{+0.25}_{-0.26} M_\oplus$, $6.05^{+0.37}_{-0.38} M_\oplus$ and $4.72^{+0.43}_{-0.42} M_\oplus$ respectively. We find the atmospheric mass fractions for TOI-270 c and d are $0.020^{+0.007}_{-0.006}$ and $0.015^{+0.007}_{-0.006}$ respectively. Meanwhile, no constraint can be placed for the initial atmospheric mass fraction of TOI-270 b since its mass is so small and its orbital period so short that there is no photoevaporative history that results in the planet not being stripped after Gyrs of evolution. Note that the adopted physical model does not include self-gravity, meaning that we cannot model large atmospheric mass fractions $X \gg 1$. As a result, we cannot place a large upper limit on TOI-270 b's initial atmospheric mass fraction, although it is unlikely to have started so high. These constraints are consistent with that of \citet{VanEylen2021}, in which the photoevaporation \verb|evapmass| code \citep{OwenEstrada2020} was used to place lower limits on the masses of TOI-270 c and d. These lower limits were found to be $1.04 M_\oplus$ and $0.44 M_\oplus$.

TOI-270 is a promising target for follow-up transit spectroscopy using HST and JWST \citep{Chouqar2020}. If the innermost planet is found to be consistent with a very small and potentially metal-rich atmosphere, it would further attest to the atmospheric mass-loss scenario since it is the H/He material that is most prone to escape, leaving behind a metal-rich, low-mass atmosphere. Likewise, the presence of a H/He dominated atmosphere for TOI-270 c and d would confirm this picture.

\subsection{Atmospheric retention post disc-dispersal} \label{sec:BHMDiscuss}
In the inference analysis of Section \ref{sec:BHM}, we find that the atmospheric mass fraction, $X \equiv M_\text{atm} / M_\text{core}$, that is retained post-disc dispersal for small, close-in exoplanets approximately follows the relation:
\begin{equation}
    X \approx 0.01 \bigg ( \frac{M_\text{core}}{M_\oplus} \bigg )^{0.34}.
\end{equation}
We also infer an intrinsic scatter in this relation, implying $\sim 0.12$ dex spread in these atmospheric masses for a given core mass. It is interesting to note that in the work of \citet{Lee2015}, an analytic scaling relation was derived for the atmospheric mass fraction accrued through gaseous core accretion. As used in \cite{Jankovic2018}, this scaling relation is adapted for varying gas surface density \citep{Lee2018,FungLee2018} as follows:
\begin{equation} \label{eq:coreacc}
    \begin{split}
    X(t) \approx & \; 0.014 \; \bigg( \frac{M_\text{core}}{M_\oplus} \bigg)^{1.6} \bigg( \frac{t}{1\text{ Myr}} \bigg)^{0.4} \; \bigg( \frac{0.02}{Z} \bigg)^{0.4}
    \; \bigg( \frac{\mu}{2.37} \bigg)^{3.3} \\
    & \times  \; \bigg( \frac{1600\text{ K}}{T_\text{rcb}} \bigg)^{1.9} \; \bigg( \frac{f_\Sigma}{0.1} \bigg)^{0.12}
    \end{split}
\end{equation}
where $Z$ is atmospheric metallicity, $\mu$ is mean molecular weight, $T_\text{rcb}$ is the temperature at the radiative-convective boundary inside the atmosphere and $f_\Sigma$ is the ratio of gas surface density to that of the minimum mass solar nebula \citep{Hayashi1981}. Note that the power-law index for core mass is $1.6$, which is shown in Figure \ref{fig:XinitXfinal} with all other nominal values taken from Eq. \ref{eq:coreacc}. One can clearly see that this relation is inconsistent with the relation inferred in this work. As argued in \citet{Jankovic2018,Rogers2021}, this discrepancy can be resolved by invoking another mass-loss process, separate from XUV photoevaporation or core-powered mass-loss, that occurs during disc dispersal. As shown in evolution models from \citet{Ikomi2012,Owen2016}, protoplanetary disc dispersal can cause dramatic atmospheric escape through a process referred to as ``boil-off''. Whilst discs survive for $3-10$ Myrs, they disperse on much shorter timescales $\sim 10^5$ yrs \citep{Kenyon1995,Luhman2010,Ercolano2011,Luhman2010,Koepferl2013}. As a result, planets immersed in these nascent discs cannot remain in hydrostatic equilibrium and lose mass via a powerful hydrodynamic outflow. Using analytic arguments, \citet{Ginzburg2016} predicted a scaling relation for the atmospheric mass retained after boil-off (referred to as `spontaneous mass-loss' in this latter study) as follows:
\begin{equation} \label{eq:boiloff}
    X \approx 0.01 \; \bigg( \frac{M_\text{core}}{M_\oplus} \bigg)^{0.44} 
    \; \bigg( \frac{T_\text{eq}}{1000\text{ K}} \bigg)^{0.25} \; \bigg( \frac{t_\text{disc}}{1 \text{ Myr}} \bigg)^{0.5},
\end{equation}
where $T_\text{eq}$ is the planetary equilibrium temperature and $t_\text{disc}$ is the disc lifetime. This relation is also shown in Figure \ref{fig:XinitXfinal} and is consistent with the relationship that we infer in this work.  This suggests that despite large atmospheric masses being accreted whilst the disc survives, the boil-off phase causes a reduction in this mass that is consistent with the masses that photoevaporation can remove. Chronologically speaking, planets might accrete such that $X \propto M_\text{c}^{1.6}$ (Eq. \ref{eq:coreacc}), boil-off such that $X \propto M_\text{c}^{0.4}$ (Eq. \ref{eq:boiloff}) and then lose mass via photoevaporation and/or core-powered mass-loss to $X \propto M_\text{c}^{1.3}$ (Eq. \ref{eq:PEscaling}). The inferred spread in atmospheric mass retention may be explained by invoking a range in equilibrium temperatures and disc lifetimes from Eq. \ref{eq:boiloff}. Note that the Bayesian hierarchical model we present here can be adapted to find such trends, for example, by determining the correlation between $X_\oplus$, $\gamma$ and $\sigma_X$ with incident stellar flux. We leave this for future work. 

Of course, there are alternative solutions to the discrepancy between core accretion and atmospheric escape. Possible mechanisms to resolve this tension also include forming the planets at the very end of the disc lifetime \citep{Ikomi2012,Lee2015}, or alternatively improving the accuracy of gas accretion models to include the effects of giant mergers \citep{Liu2015,Inamdar2016} which would result in potentially significant atmospheric mass-loss and the production of heat which would take typically kyrs to disperse. The inclusion of 3D simulations has additionally shown that recycling of high-entropy gas during the accretion phase can act to reduce the final atmospheric mass of the planet \citep{Ormel2015,Fung2015,Cimerman2017,Ali-Dib2020,Chen2020}. We highlight that more sophisticated modelling of the boil-off phase is required, as it is clearly an important process in exoplanet evolution to understand.

\subsection{The curious case of Kepler 36 c} \label{sec:Kepler36c}
Kepler 36 c is the only planet for which its mass and size are not accurately reproduced by the BHM model (see Figure \ref{fig:Latent}). In \citet{OwenMorton2016}, a similar analysis to that performed on TOI-270 in Section \ref{sec:TOI-270} was performed on this system to find that Kepler 36 c required an initial atmospheric mass fraction of $\sim 20\%$ under the photoevaporative model (similar to the early suggestion of \citealt{LopezFortney2013} for this system), which is much higher than predicted by our statistical relation between core mass and initial atmospheric mass fraction. This is likely caused by the fact that Kepler 36 c has one of the largest sizes of all planets in our sample ($R_\text{p} \approx 3.7R_\oplus$) given its mass ($M_\text{p} \approx 7.1M_\oplus$) \citep{Carter2012,Vissapragada2020}. Furthermore, the planet is in a delicate 7:6 mean-motion resonance with Kepler 36 b, hinting towards a dynamical history with potential mutual migration \citep{Quillen2013,Rimlinger2021}, and perhaps a smoother than normal disc dispersal process. One can conclude that Kepler 36 c is highly inflated and may not be well-described by our statistical model. {\jr{Investigating how other highly inflated planets \citep[e.g.][]{Bonfils2012,Ofir2014,JontofHutter2014,Wang2019,Belkovski2022} fit within our statistical model if left for future work, however we note that spectroscopic follow-up of Kepler 36 c as well as other similarly inflated planets may help guide our interpretation of this result.}}

\subsection{Core Compositions} \label{sec:coreCompsDiscuss}
The inferred core compositions from the inference model of Section \ref{sec:BHM} have a mean of $\mu_\rho = 0.17^{+0.07}_{-0.08}$, which implies an iron-silicate mixture with iron-mass-fraction of $17\%$. For an Earth-mass core, this would result in a bulk density of $\rho_{M_\oplus} \approx 4.7 \text{ g cm}^{-3}$. This is consistent with compositions determined for the TOI-270 planets from Section \ref{sec:TOI-270} as well as the inference analysis of \citet{Rogers2021}, which inferred a value of $0.26^{+0.08}_{-0.09}$ for the sample of planets from the \textit{California Kepler Survey} (CKS) \citep{Fulton2017}. In this latter work, orbital periods and planet sizes were used to infer the underlying core mass distribution, initial atmospheric mass fraction and core composition distribution. To do so, synthetic transit surveys were implemented to accurately model the bias of the \textit{Kepler} survey. As a result, the constraints placed on the distributions of interest were representative of the underlying planet distribution. In the inference model we present in this paper, the distribution for core composition is not representative since the adopted planet sample does not have quantifiable biases. Instead, this distribution represents the compositions for the cores of the adopted sample with measured masses and radii. Note that RV/TTV surveys are heavily biased to observe larger mass planets, yet this bias is extremely difficult to quantify since it would require homogeneous observations of a well-defined sample of stars \citep{Howard2010,Mayor2011,Fulton2021}. Nevertheless, the inferred core compositions are inconsistent with significant ice-mass fractions and thus hint towards core formation interior to water-ice line \citep[e.g.][]{Hansen2012,Chiang2013,Chatterjee2014,Jankovic2018}. Recall, however, that our model assumes all planets in a given system have the same core composition. This would not be the case for models that posit that super-Earths form interior to the water-ice line and sub-Neptunes are in fact water-rich planets that migrated into their current orbital separation \citep[e.g.][]{Ida2005,Ida2010,Bodenheimer2014,Raymond2014,Bitsch2018b, Raymond2018, Zeng2019}. Of particular note, \citet{Neil2022} explored the statistical validity of such `water worlds' within the framework of a joint mass-radius-period distribution for a selection of \textit{Kepler} planets. They compared various population mixture models consisting of different combinations of planet types; including rocky and icy worlds, both of which could host H/He atmospheres or be stripped/born intrinsically rocky/icy. They found strong degeneracy between water-world populations and rocky planets that host atmospheres, suggesting that the current data, specifically sub-Neptunes, are consistent with both scenarios. We, on other hand, find more favour for rocky cores with H/He atmospheres, as opposed to water worlds. As mentioned, this comes from the fact that we assume all cores in a system have the same core composition and that, unlike \citet{Neil2022}, we explicitly perform evolution calculations on the planets in our models. We are thus directly exploiting the evolution of each planet to find the histories that are consistent with its current mass and radius.

Whilst our value for mean composition is consistent with \citet{Rogers2021}, we infer $\sim 10\%$ less iron content in the cores of planets. There are multiple potential causes to such a difference. One possibility is the differing planet samples. In \citet{Rogers2021}, the CKS sample was used, which exclusively included FGK stars \citep{CKSI-Petigura2017}. In this work, however, we have included stars with masses down to $0.3 M_\odot$. Since different formation mechanisms may be at play around lower-mass stars, the addition of planets orbiting M-dwarfs could reduce the inferred mean core composition, implying that lower-mass stars host less iron-rich cores. The Bayesian hierarchical framework presented in Section \ref{sec:BHM} is suitable to investigate such trends, for example, by determining the scaling between stellar mass and core composition. Another option, however comes from the works of \citet{Kite2016,Schlichting2022}, in which chemical modelling of planetary cores interacting with H/He dominated atmospheres demonstrated significant sequestration of hydrogen and oxygen into the metal cores, reducing its bulk density when compared to Earth. \citet{Schlichting2022} argue that this may be the cause for the observed reduction in bulk densities of Trappist-1 planets \citep[e.g.][]{Dorn2018,Dorn2019,Agol2021}.

Finally, another cause may come from the different methodology of the work we present here and that of \citet{Rogers2021}. In the latter, the position of the radius gap itself allows constraints to be placed on the core compositions of \textit{Kepler} planets. As shown in \citet{Owen2017}, the radius gap shifts to smaller radii for populations of planets with an iron-rich core composition since this compresses a core for a given mass. In the work we present here, we are not directly using the radius gap as a demographic feature, but instead using the measured masses and radii from \textit{Kepler}, \textit{K2} and \textit{TESS} planets. This may also explain the difference in core composition standard deviation, inferred to be $\sigma_\rho = 0.10^{+0.03}_{-0.03}$ in this work and an upper limit of $\sigma_\rho \leq 0.16$ in \citet{Rogers2021}. Whilst still statistically consistent, this difference may be caused by the varying approaches. On a demographic level, a clean radius gap is produced by populations of planets with a small spread in core compositions. Thus, it is clear why the inference analysis of \citet{Rogers2021} inferred a value consistent with no spread. Ideally, one would construct a combined approach, which uses large demographic distributions of planets around the radius gap as well as measured masses to fully constrain their core compositions. 

\subsection{Atmospheric evolution emulators} \label{sec:emulatorDiscuss}
The results shown in Figures \ref{fig:TOI270posteriors} and \ref{fig:TOI270composition} clearly demonstrate that the evolution emulator is capable of reproducing the demographics and constraints provided by the semi-analytic model of \citet{Owen2017}, whilst crucially being dramatically quicker to compute. The goal of this paper was to validate the use of such an emulator, paving the way for more complex models to be used as training data. We note that the work of \citet{OwenMorton2016}, which placed constraints on the initial conditions of Kepler-36 b and c \citep{Carter2012}, implemented evolutionary models of photoevaporation using the stellar and planetary evolution code \verb|mesa| \citep{Paxton2011,Paxton2013,Paxton2015,Paxton2018}. Whilst being more accurate than the semi-analytic model adopted for training in this work, the computational expense was far greater since a grid of simulations was produced and interpolated in the MCMC sampling. 

The adopted semi-analytic model of the planetary structure is known to perform poorly when the atmospheric mass fraction falls to small values, such that the radiative zone of the atmosphere starts to dominate both the radius and envelope mass. Since the evolution emulator can now be trained on more complex simulations that do not suffer from such issues, one could construct and train the emulator on a suite of planetary evolution simulations that incorporate appropriate radiative transfer, equations of state and self-gravity \citep[e.g., those determined by MESA][]{Owen2013,ChenRogers2016,Kubyshkina2020}. One could also incorporate additional physics from alternative mass-loss models such as core-powered mass-loss \citep[e.g.][]{Ginzburg2018,Gupta2019,Rogers2021b} to investigate how the two mechanisms compete. Furthermore, it is important to emphasise the issue of `over-fitting' is not of concern in this machine learning application since the deterministic models do not introduce noise that the emulator may be unintentionally trained on. This is because a single set of planetary and stellar conditions always produce the same planetary evolution track. {\jr{Moreover, our training data samples parameter space to a high degree and crucially in a random manner (as opposed to grid-based sampling), meaning that over-fitting does not become a further issue.}}

In this work, the emulator was trained with input parameters of orbital period, stellar mass, stellar metallicity, core mass, core density and initial atmospheric mass fraction. As a further improvement, one could include additional parameters such as the planet's initial cooling timescale and the system's final age. We note that this latter variable has little effect on the final radius of a planet under the photoevaporation model since the majority of mass-loss occurs during the first $\sim 100$Myrs. As the majority of observed exoplanets orbit main-sequence stars, it implies that age is not a dominant parameter. However, the small fraction of planets that the emulator misclassifies are those that are being stripped on Gyr timescales (as shown in Section \ref{sec:emulatorDesign}). Therefore, adding age as a parameter {\jr{may}} reduce this error since the emulator {\jr{could}} learn this trend. Furthermore, if additional physics is included, such as core-powered mass-loss, which operates at much longer timescales, one would also require system age as an input.

\section{Conclusions}
We have shown that an emulator, trained on an evolution model for an exoplanet's H/He atmosphere, can accurately predict the final properties of an exoplanet at a fraction of the computational expense of standard evolutionary models. Given that exoplanet evolution results in a bimodal population of super-Earths and sub-Neptunes, we implemented an emulator that consists of two neural networks: the first that classifies planets that will be stripped of their primordial H/He atmosphere from those that won't; and a second that determines the final planet size and atmospheric mass fraction for those that can maintain such an atmosphere after $5$ Gyrs. We find that the fractional RMS error introduced in the final radius is $\sim 1\%$ when compared to the original model, typically smaller than even the best measurements of an exoplanet's radius \citep[for example using asteroseismology, e.g.][]{VanEylen2018}. The computational speed-up factor is $\sim 1000$.

As a test-case, we use the evolution emulator to infer the initial conditions of TOI-270 b, c and d \citep{VanEylen2021}. We find that the inner planet is stripped, whilst the outer planets have maintained a H/He atmosphere which was originally $1-2\%$ before photoevaporation took place. These constraints were near-indistinguishable from those found when using the original semi-analytic photoevaporation model from \citet{Owen2017}, validating the use of the emulator.

We then applied the evolution emulator to a more sophisticated inference problem of determining the initial atmospheric conditions for a sample of 81 planets with well-constrained masses and radii from \textit{Kepler}, \textit{K2} and \textit{TESS}. In doing so, we inferred the relation between core mass and the atmospheric mass fraction retained post-protoplanetary disc dispersal. The main conclusions from this analysis are as follows:

\begin{itemize}
    \setlength\itemsep{1em}
    \item We find that small, close-in planets retain an atmospheric mass fraction after disc dispersal, $X \equiv M_\text{atm} / M_\text{core}$, according to $X \approx 0.01 ( M_\text{core} / {M_\oplus})^{0.34}$, which is consistent with the analytic predictions from \citet{Ginzburg2016} that include the physics of mass-loss during disc dispersal, often referred to as the boil-off phase \citep{Owen2016}. We also find an intrinsic scatter to these retained atmospheric masses of $\sim 0.12$ dex, which may arise due to differing disc lifetimes and planet equilibrium temperatures. We highlight that more sophisticated modelling of atmospheric escape during disc dispersal is required to understand its importance in exoplanet evolution better.
    
    \item The mean core composition for the planet sample was found to be an iron-silicate mixture, with an iron-mass-fraction of $0.17^{+0.07}_{-0.08}$. For an Earth-mass core, this would result in a bulk density of $\rho_{M_\oplus} \approx 4.7 \text{ g cm}^{-3}$. Whilst this is consistent with the inference analysis of \citet{Rogers2021}, which was applied to the \textit{California Kepler Survey} \citet{Fulton2017}, we note the value inferred here is lower in iron-mass-fraction. This could arise due to differing methodologies or potentially due to a different planet sample, including planets around lower mass stars. It could also suggest an increase in hydrogen and oxygen sequestration into the metal cores for such planets \citep{Schlichting2022}.
    
\end{itemize}

As with \citet{Rogers2021}, these results depend on the fact that the photoevaporation scenario dominates the evolution of small, close-in exoplanets. We highlight that more theoretical work is needed to understand the interplay between photoevaporation and core-powered mass-loss \citep{Rogers2021b,Schulik2022}, as well as other escape mechanisms such as the boil-off phase, which acts during protoplanetary disc dispersal. 

Overall, this work has demonstrated that machine-learnt emulators are well-suited for demographic inference analyses. Although the emulator is clearly accurate and fast, there are many aspects in which it may be improved. Instead of simple evolutionary models, sophisticated numerical simulations with higher accuracy can now be used for training. Furthermore, other parameters, such as system age and initial cooling timescale, may be used as inputs. Finally, additional physics may be included in the training data, such as core-powered mass-loss \citep{Ginzburg2018,Gupta2019}. Nevertheless, implementing an appropriate evolution emulator to inference analyses will remedy the issues of computational cost and thus allow investigation into potential correlations between distributions such as stellar mass and core composition, as well as the imprint of competing mass-loss mechanisms on the demographics of exoplanets.

\section*{Acknowledgements}

We are grateful to the anonymous referee for comments which improved the manuscript. We kindly thank Richard Booth for comments and discussion that helped improve the paper. JGR is supported by a 2017 Royal Society Grant for Research Fellows. CJM was supported by a 2020 Royal Society Enhancement Award. JEO is supported by a Royal Society University Research Fellowship. This work was supported by the European Research Council (ERC) under the European Union’s Horizon 2020 research and innovation programme (Grant agreement No. 853022, PEVAP). This research has made use of the NASA Exoplanet Archive, which is operated by the California Institute of Technology, under contract with the National Aeronautics and Space Administration under the Exoplanet Exploration Program. This work was performed using the DiRAC Data Intensive service at Leicester, operated by the University of Leicester IT Services, which forms part of the STFC DiRAC HPC Facility (www.dirac.ac.uk). The equipment was funded by BEIS capital funding via STFC capital grants ST/K000373/1 and ST/R002363/1 and STFC DiRAC Operations grant ST/R001014/1. DiRAC is part of the National e-Infrastructure. JGR and JEO are grateful for hospitality from UCLA, where this work was initiated. 

\section*{Data Availability}
The MCMC chains for the population-level Bayesian Hierarchical model, as well as a machine-readable version of Table \ref{tab:models} are made available at \href{https://doi.org/10.5281/zenodo.7150445}{10.5281/zenodo.7150445}.



\bibliographystyle{mnras}
\bibliography{references} 



\appendix

\begin{table*}
	\centering
	\def\arraystretch{1.5}
	\caption{Planetary properties from our evolutionary inference model.}
	\begin{tabular}{ lcccr } 
		\hline
		\multicolumn{1}{|p{2.5cm}|}{\centering Planet \\ Name} & \multicolumn{1}{|p{2.5cm}|}{\centering Core \\ Mass ($M_\oplus$)} & \multicolumn{1}{|p{2.5cm}|}{\centering Initial Atmospheric \\ Mass Fraction $(\%)$} & \multicolumn{1}{|p{2.5cm}|}{\centering Final Atmospheric \\ Mass Fraction $(\%)$} & \multicolumn{1}{|p{2.5cm}|}{\centering Parameter \\ Reference} \\
		\hline\hline
        EPIC 249893012 b & $7.40^{{}+0.74{}}_{{}-0.70{}}$ & $1.58^{{}+0.91{}}_{{}-0.71{}}$ & $\leq0.19$ & \citet{Hidalgo2020} \\
        EPIC 249893012 c & $15.14^{{}+2.18{}}_{{}-1.70{}}$ & $5.02^{{}+1.29{}}_{{}-1.29{}}$ & $4.74^{{}+1.15{}}_{{}-1.17{}}$ & \citet{Hidalgo2020} \\
        GJ 3470 b & $12.99^{{}+1.58{}}_{{}-1.55{}}$ & $3.35^{{}+1.95{}}_{{}-1.05{}}$ & $3.07^{{}+1.46{}}_{{}-0.95{}}$ & \citet{Kosiarek2019} \\
        GJ 357 b & $1.85^{{}+0.32{}}_{{}-0.33{}}$ & $1.19^{{}+0.73{}}_{{}-0.47{}}$ & $\leq0.01$ & \citet{Luque2019} \\
        GJ 486 b & $2.80^{{}+0.17{}}_{{}-0.15{}}$ & $1.47^{{}+0.77{}}_{{}-0.56{}}$ & $\leq0.01$ & \citet{Trifonov2021} \\
        GJ 9827 b & $4.91^{{}+0.43{}}_{{}-0.52{}}$ & $1.73^{{}+0.88{}}_{{}-0.66{}}$ & $\leq0.01$ & \citet{Dai2019} \\
        HD 110113 b & $5.11^{{}+0.68{}}_{{}-0.60{}}$ & $1.56^{{}+0.95{}}_{{}-0.54{}}$ & $\leq0.01$ & \citet{Osborn2021} \\
        HD 136352 b & $3.75^{{}+0.31{}}_{{}-0.36{}}$ & $1.16^{{}+0.62{}}_{{}-0.44{}}$ & $\leq0.01$ & \citet{Kane2020} \\
        HD 136352 c & $10.74^{{}+0.88{}}_{{}-1.01{}}$ & $1.99^{{}+0.61{}}_{{}-0.56{}}$ & $1.94^{{}+0.59{}}_{{}-0.53{}}$ & \citet{Kane2020} \\
        HD 137496 b & $3.41^{{}+0.57{}}_{{}-0.53{}}$ & $1.64^{{}+0.77{}}_{{}-0.65{}}$ & $\leq0.01$ & \citet{AzevedoSilva2022} \\
        HD 15337 b & $5.40^{{}+0.56{}}_{{}-0.60{}}$ & $1.05^{{}+0.72{}}_{{}-0.39{}}$ & $\leq0.01$ & \citet{Dumusque2019} \\
        HD 15337 c & $8.97^{{}+1.58{}}_{{}-1.45{}}$ & $1.96^{{}+0.62{}}_{{}-0.60{}}$ & $1.85^{{}+0.55{}}_{{}-0.56{}}$ & \citet{Dumusque2019} \\
        HD 191939 b & $10.93^{{}+1.34{}}_{{}-1.11{}}$ & $6.16^{{}+1.09{}}_{{}-0.85{}}$ & $4.79^{{}+0.64{}}_{{}-0.52{}}$ & \citet{Lubin2022} \\
        HD 191939 c & $8.84^{{}+1.87{}}_{{}-1.24{}}$ & $4.42^{{}+1.10{}}_{{}-0.88{}}$ & $4.21^{{}+0.92{}}_{{}-0.84{}}$ & \citet{Lubin2022} \\
        HD 207897 b & $13.20^{{}+2.03{}}_{{}-1.98{}}$ & $1.51^{{}+0.55{}}_{{}-0.43{}}$ & $1.49^{{}+0.53{}}_{{}-0.42{}}$ & \citet{Heidari2022} \\
        HD 213885 b & $8.64^{{}+0.80{}}_{{}-0.75{}}$ & $2.15^{{}+1.04{}}_{{}-0.76{}}$ & $\leq0.01$ & \citet{Espinoza2020} \\
        HD 219134 b & $4.74^{{}+0.26{}}_{{}-0.29{}}$ & $1.65^{{}+0.88{}}_{{}-0.61{}}$ & $\leq0.01$ & \citet{Gillon2017} \\
        HD 219134 c & $3.99^{{}+0.27{}}_{{}-0.23{}}$ & $0.94^{{}+0.38{}}_{{}-0.28{}}$ & $\leq0.01$ & \citet{Gillon2017} \\
        HD 219134 d & $10.67^{{}+0.00{}}_{{}-0.00{}}$ & $0.49^{{}+0.00{}}_{{}-0.00{}}$ & $\leq0.01$ & \citet{Gillon2017} \\
        HD 260655 b & $1.96^{{}+0.27{}}_{{}-0.25{}}$ & $1.21^{{}+0.80{}}_{{}-0.48{}}$ & $\leq0.01$ & \citet{Luque2022} \\
        HD 260655 c & $2.83^{{}+0.31{}}_{{}-0.33{}}$ & $1.01^{{}+0.55{}}_{{}-0.37{}}$ & $\leq0.17$ & \citet{Luque2022} \\
        HD 5278 b & $7.63^{{}+1.14{}}_{{}-0.91{}}$ & $1.81^{{}+0.68{}}_{{}-0.54{}}$ & $1.15^{{}+0.36{}}_{{}-0.34{}}$ & \citet{Sozzetti2021} \\
        HD 63935 b & $10.74^{{}+1.75{}}_{{}-1.57{}}$ & $2.86^{{}+0.97{}}_{{}-0.79{}}$ & $2.39^{{}+0.76{}}_{{}-0.62{}}$ & \citet{Scarsdale2021} \\
        HD 73583 c & $8.96^{{}+1.91{}}_{{}-1.62{}}$ & $1.65^{{}+0.62{}}_{{}-0.50{}}$ & $1.61^{{}+0.59{}}_{{}-0.48{}}$ & \citet{Barragan2022} \\
        HD 86226 c & $7.58^{{}+0.64{}}_{{}-0.63{}}$ & $1.96^{{}+1.02{}}_{{}-0.70{}}$ & $0.23^{{}+0.23{}}_{{}-0.14{}}$ & \citet{Teske2020} \\
        HD 97658 b & $7.17^{{}+0.96{}}_{{}-1.00{}}$ & $1.53^{{}+0.59{}}_{{}-0.51{}}$ & $1.15^{{}+0.42{}}_{{}-0.40{}}$ & \citet{VanGrootel2014} \\
        K2-110 b & $13.20^{{}+2.99{}}_{{}-2.79{}}$ & $1.77^{{}+0.67{}}_{{}-0.54{}}$ & $1.72^{{}+0.63{}}_{{}-0.53{}}$ & \citet{Osborn2017} \\
        K2-111 b & $5.47^{{}+0.47{}}_{{}-0.58{}}$ & $1.55^{{}+0.91{}}_{{}-0.59{}}$ & $\leq0.11$ & \citet{Mortier2020} \\
        K2-146 b & $5.90^{{}+0.65{}}_{{}-0.63{}}$ & $1.95^{{}+0.80{}}_{{}-0.59{}}$ & $0.95^{{}+0.39{}}_{{}-0.37{}}$ & \citet{Lam2020} \\
        K2-18 b & $8.97^{{}+1.58{}}_{{}-1.86{}}$ & $2.00^{{}+0.89{}}_{{}-0.68{}}$ & $2.00^{{}+0.91{}}_{{}-0.69{}}$ & \citet{Sarkis2018} \\
        K2-180 b & $10.24^{{}+2.06{}}_{{}-1.94{}}$ & $1.38^{{}+0.52{}}_{{}-0.51{}}$ & $1.20^{{}+0.42{}}_{{}-0.45{}}$ & \citet{Korth2019} \\
        K2-199 c & $12.24^{{}+2.25{}}_{{}-2.01{}}$ & $2.58^{{}+0.88{}}_{{}-0.65{}}$ & $2.43^{{}+0.74{}}_{{}-0.60{}}$ & \citet{AkanaMurphy2021} \\
        K2-285 b & $10.02^{{}+1.13{}}_{{}-0.97{}}$ & $2.07^{{}+0.79{}}_{{}-0.59{}}$ & $1.16^{{}+0.33{}}_{{}-0.36{}}$ & \citet{Palle2019} \\
        K2-285 c & $16.50^{{}+1.61{}}_{{}-2.16{}}$ & $5.06^{{}+0.97{}}_{{}-0.81{}}$ & $4.65^{{}+0.61{}}_{{}-0.72{}}$ & \citet{Palle2019} \\
        K2-291 b & $5.26^{{}+0.87{}}_{{}-0.68{}}$ & $1.78^{{}+0.96{}}_{{}-0.65{}}$ & $\leq0.01$ & \citet{Dai2019} \\
        KOI-142 b & $10.89^{{}+1.19{}}_{{}-1.45{}}$ & $5.92^{{}+1.68{}}_{{}-0.88{}}$ & $4.52^{{}+0.68{}}_{{}-0.56{}}$ & \citet{Weiss2020} \\
        Kepler-10 c & $11.95^{{}+1.97{}}_{{}-2.06{}}$ & $1.27^{{}+0.46{}}_{{}-0.40{}}$ & $1.26^{{}+0.46{}}_{{}-0.39{}}$ & \citet{Weiss2016} \\
        Kepler-107 c & $5.50^{{}+0.63{}}_{{}-0.78{}}$ & $1.84^{{}+0.88{}}_{{}-0.71{}}$ & $\leq0.01$ & \citet{Bonomo2019} \\
        Kepler-1705 b & $4.92^{{}+0.50{}}_{{}-0.48{}}$ & $1.87^{{}+0.86{}}_{{}-0.71{}}$ & $\leq0.29$ & \citet{Leleu2021b} \\
        Kepler-1705 c & $5.46^{{}+0.58{}}_{{}-0.60{}}$ & $1.80^{{}+0.96{}}_{{}-0.61{}}$ & $0.29^{{}+0.29{}}_{{}-0.20{}}$ & \citet{Leleu2021b} \\
        Kepler-177 b & $6.22^{{}+1.25{}}_{{}-1.02{}}$ & $6.65^{{}+2.20{}}_{{}-2.79{}}$ & $5.46^{{}+1.76{}}_{{}-1.94{}}$ & \citet{Vissapragada2020} \\        
        \end{tabular}
	\label{tab:models}
\end{table*}

\begin{table*}
	\centering
	\def\arraystretch{1.5}
	\begin{tabular}{ lcccr } 
		\hline
		\multicolumn{1}{|p{2.5cm}|}{\centering Planet \\ Name} & \multicolumn{1}{|p{2.5cm}|}{\centering Core \\ Mass ($M_\oplus$)} & \multicolumn{1}{|p{2.5cm}|}{\centering Initial Atmospheric \\ Mass Fraction $(\%)$} & \multicolumn{1}{|p{2.5cm}|}{\centering Final Atmospheric \\ Mass Fraction $(\%)$} & \multicolumn{1}{|p{2.5cm}|}{\centering Parameter \\ Reference} \\
		\hline\hline
        Kepler-26 b & $5.48^{{}+0.66{}}_{{}-0.64{}}$ & $3.47^{{}+1.46{}}_{{}-0.87{}}$ & $2.96^{{}+0.98{}}_{{}-0.69{}}$ & \citet{JontofHutter2016} \\
        Kepler-26 c & $6.68^{{}+0.74{}}_{{}-0.75{}}$ & $2.85^{{}+1.09{}}_{{}-0.90{}}$ & $2.75^{{}+0.99{}}_{{}-0.87{}}$ & \citet{JontofHutter2016} \\
        Kepler-30 b & $11.78^{{}+1.71{}}_{{}-1.30{}}$ & $6.50^{{}+1.90{}}_{{}-1.87{}}$ & $6.33^{{}+1.66{}}_{{}-1.85{}}$ & \citet{SanchisOjeda2012} \\
        Kepler-307 b & $7.09^{{}+1.07{}}_{{}-0.84{}}$ & $1.83^{{}+0.70{}}_{{}-0.55{}}$ & $1.37^{{}+0.48{}}_{{}-0.41{}}$ & \citet{JontofHutter2016} \\
        Kepler-307 c & $4.42^{{}+0.51{}}_{{}-0.51{}}$ & $1.89^{{}+0.79{}}_{{}-0.57{}}$ & $0.96^{{}+0.37{}}_{{}-0.34{}}$ & \citet{JontofHutter2016} \\
        Kepler-36 b & $3.78^{{}+0.13{}}_{{}-0.15{}}$ & $1.14^{{}+0.48{}}_{{}-0.38{}}$ & $\leq0.01$ & \citet{Vissapragada2020} \\
        Kepler-36 c & $7.40^{{}+0.00{}}_{{}-0.41{}}$ & $3.62^{{}+2.36{}}_{{}-0.00{}}$ & $2.32^{{}+0.63{}}_{{}-0.00{}}$ & \citet{Vissapragada2020} \\
        Kepler-48 c & $13.42^{{}+2.60{}}_{{}-2.33{}}$ & $2.08^{{}+0.80{}}_{{}-0.62{}}$ & $1.97^{{}+0.72{}}_{{}-0.59{}}$ & \citet{Marcy2014} \\
        Kepler-60 b & $4.46^{{}+0.57{}}_{{}-0.57{}}$ & $1.56^{{}+0.89{}}_{{}-0.61{}}$ & $\leq0.01$ & \citet{JontofHutter2016} \\
        Kepler-80 b & $7.47^{{}+0.81{}}_{{}-0.85{}}$ & $2.60^{{}+0.90{}}_{{}-0.78{}}$ & $1.89^{{}+0.56{}}_{{}-0.56{}}$ & \citet{Macdonald2016} \\
        Kepler-80 c & $7.51^{{}+0.91{}}_{{}-0.97{}}$ & $2.79^{{}+0.96{}}_{{}-0.78{}}$ & $2.31^{{}+0.71{}}_{{}-0.64{}}$ & \citet{Macdonald2016} \\
        Kepler-80 d & $5.37^{{}+0.51{}}_{{}-0.55{}}$ & $1.25^{{}+0.87{}}_{{}-0.47{}}$ & $\leq0.01$ & \citet{Macdonald2016} \\
        Kepler-80 e & $3.87^{{}+0.46{}}_{{}-0.50{}}$ & $1.32^{{}+0.70{}}_{{}-0.50{}}$ & $\leq0.01$ & \citet{Macdonald2016} \\
        Kepler-93 b & $4.65^{{}+0.83{}}_{{}-0.86{}}$ & $1.62^{{}+0.91{}}_{{}-0.61{}}$ & $\leq0.01$ & \citet{Stassun2017} \\
        Kepler-94 b & $13.10^{{}+1.46{}}_{{}-1.22{}}$ & $4.00^{{}+2.38{}}_{{}-1.13{}}$ & $2.62^{{}+0.89{}}_{{}-0.76{}}$ & \citet{Marcy2014} \\
        L 168-9 b & $4.34^{{}+0.64{}}_{{}-0.70{}}$ & $1.65^{{}+0.92{}}_{{}-0.58{}}$ & $\leq0.01$ & \citet{AstudilloDefru2020} \\
        L 98-59 c & $2.29^{{}+0.30{}}_{{}-0.30{}}$ & $1.23^{{}+0.76{}}_{{}-0.48{}}$ & $\leq0.01$ & \citet{Demangeon2021} \\
        L 98-59 d & $2.05^{{}+0.27{}}_{{}-0.23{}}$ & $1.28^{{}+0.68{}}_{{}-0.43{}}$ & $\leq0.42$ & \citet{Demangeon2021} \\
        TOI-1062 b & $9.23^{{}+0.94{}}_{{}-1.09{}}$ & $1.44^{{}+0.63{}}_{{}-0.50{}}$ & $0.75^{{}+0.33{}}_{{}-0.29{}}$ & \citet{Otegi2021} \\
        TOI-1064 b & $12.23^{{}+1.84{}}_{{}-2.03{}}$ & $1.64^{{}+0.57{}}_{{}-0.44{}}$ & $1.54^{{}+0.48{}}_{{}-0.40{}}$ & \citet{Wilson2022} \\
        TOI-1235 b & $4.66^{{}+0.45{}}_{{}-0.48{}}$ & $1.03^{{}+0.57{}}_{{}-0.47{}}$ & $\leq0.24$ & \citet{Bluhm2020} \\
        TOI-1246 b & $10.15^{{}+1.14{}}_{{}-1.07{}}$ & $3.86^{{}+2.30{}}_{{}-1.11{}}$ & $2.20^{{}+0.48{}}_{{}-0.56{}}$ & \citet{Turtelboom2022} \\
        TOI-1246 c & $8.49^{{}+1.20{}}_{{}-1.15{}}$ & $1.91^{{}+0.84{}}_{{}-0.57{}}$ & $1.30^{{}+0.42{}}_{{}-0.39{}}$ & \citet{Turtelboom2022} \\
        TOI-1246 e & $15.35^{{}+2.27{}}_{{}-1.98{}}$ & $6.08^{{}+1.81{}}_{{}-1.32{}}$ & $6.04^{{}+1.78{}}_{{}-1.33{}}$ & \citet{Turtelboom2022} \\
        TOI-125 b & $9.91^{{}+0.81{}}_{{}-0.91{}}$ & $2.56^{{}+0.95{}}_{{}-0.72{}}$ & $1.57^{{}+0.42{}}_{{}-0.42{}}$ & \citet{Nielsen2020} \\
        TOI-125 c & $7.53^{{}+0.96{}}_{{}-0.91{}}$ & $2.83^{{}+0.96{}}_{{}-0.83{}}$ & $1.93^{{}+0.52{}}_{{}-0.53{}}$ & \citet{Nielsen2020} \\
        TOI-125 d & $13.39^{{}+1.53{}}_{{}-1.46{}}$ & $2.58^{{}+0.99{}}_{{}-0.81{}}$ & $2.53^{{}+0.96{}}_{{}-0.80{}}$ & \citet{Nielsen2020} \\
        TOI-1260 b & $7.75^{{}+1.24{}}_{{}-0.97{}}$ & $1.89^{{}+0.77{}}_{{}-0.66{}}$ & $0.89^{{}+0.40{}}_{{}-0.35{}}$ & \citet{Georgieva2021} \\
        TOI-1759 b & $11.65^{{}+1.53{}}_{{}-1.51{}}$ & $4.29^{{}+1.08{}}_{{}-1.00{}}$ & $4.26^{{}+1.08{}}_{{}-0.95{}}$ & \citet{Espinoza2022} \\
        TOI-178 c & $4.65^{{}+0.50{}}_{{}-0.58{}}$ & $1.53^{{}+0.72{}}_{{}-0.56{}}$ & $\leq0.01$ & \citet{Leleu2021a} \\
        TOI-220 b & $13.64^{{}+1.32{}}_{{}-1.26{}}$ & $2.76^{{}+0.99{}}_{{}-0.77{}}$ & $2.62^{{}+0.91{}}_{{}-0.71{}}$ & \citet{Hoyer2021} \\
        TOI-269 b & $9.45^{{}+1.67{}}_{{}-1.28{}}$ & $2.71^{{}+0.84{}}_{{}-0.83{}}$ & $2.28^{{}+0.67{}}_{{}-0.66{}}$ & \citet{Cointepas2021} \\
        TOI-270 b & $1.61^{{}+0.25{}}_{{}-0.21{}}$ & $1.13^{{}+0.76{}}_{{}-0.47{}}$ & $\leq0.01$ & \citet{VanEylen2021} \\
        TOI-270 c & $6.12^{{}+0.48{}}_{{}-0.49{}}$ & $2.12^{{}+0.69{}}_{{}-0.59{}}$ & $1.70^{{}+0.49{}}_{{}-0.50{}}$ & \citet{VanEylen2021} \\
        TOI-270 d & $4.78^{{}+0.49{}}_{{}-0.56{}}$ & $1.69^{{}+0.51{}}_{{}-0.50{}}$ & $1.46^{{}+0.45{}}_{{}-0.44{}}$ & \citet{VanEylen2021} \\
        TOI-431 d & $11.08^{{}+1.90{}}_{{}-1.69{}}$ & $4.65^{{}+1.28{}}_{{}-1.06{}}$ & $4.39^{{}+1.09{}}_{{}-0.95{}}$ & \citet{Osborn2021} \\
        TOI-561 c & $6.85^{{}+0.89{}}_{{}-0.58{}}$ & $3.48^{{}+1.65{}}_{{}-0.95{}}$ & $2.27^{{}+0.66{}}_{{}-0.57{}}$ & \citet{Lacedelli2021} \\
        TOI-763 b & $8.90^{{}+1.29{}}_{{}-0.84{}}$ & $1.50^{{}+0.74{}}_{{}-0.52{}}$ & $0.94^{{}+0.37{}}_{{}-0.36{}}$ & \citet{Fridlund2020} \\
        TOI-763 c & $9.23^{{}+1.10{}}_{{}-1.32{}}$ & $2.14^{{}+0.76{}}_{{}-0.66{}}$ & $1.86^{{}+0.61{}}_{{}-0.55{}}$ & \citet{Fridlund2020} \\
        TOI-824 b & $17.25^{{}+1.56{}}_{{}-1.93{}}$ & $2.65^{{}+1.20{}}_{{}-0.94{}}$ & $1.63^{{}+0.64{}}_{{}-0.56{}}$ & \citet{Burt2020} \\
        Wolf 503 b & $5.77^{{}+0.72{}}_{{}-0.58{}}$ & $1.39^{{}+0.72{}}_{{}-0.48{}}$ & $0.52^{{}+0.29{}}_{{}-0.21{}}$ & \citet{Polanski2021} \\
        pi Men c & $5.71^{{}+0.41{}}_{{}-0.36{}}$ & $1.91^{{}+0.88{}}_{{}-0.61{}}$ & $0.20^{{}+0.17{}}_{{}-0.10{}}$ & \citet{Gandolfi2018} \\
\end{tabular}
\end{table*}


\bsp	
\label{lastpage}
\end{document}